\documentclass[traditabstract]{aa}
\usepackage{txfonts}
\usepackage{graphicx}
\usepackage{hyperref}
\usepackage{natbib}
\bibpunct{(}{)}{;}{a}{}{,} 
\usepackage{epsfig}

\newcommand{\teff}{\mbox{$T_{\rm eff}$}}
\newcommand{\logg}{\mbox{$\log g$}}
\newcommand{\vsini}{\mbox{$v \sin i_\star$ }}
\newcommand{\mictrb}{\mbox{$\xi_{\rm t}$}}
\newcommand{\mactrb}{\mbox{$v_{\rm mac}$}}

\newcommand{\halpha}{\mbox{$H_\alpha$}}

\begin{document}
\title{The EBLM Project\thanks{using WASP-South photometric observations (Sutherland, South Africa) confirmed with radial velocity measurement from the CORALIE spectrograph, photometry from the \textit{EulerCam} camera (both mounted on the Swiss 1.2\,m \textit{Euler} Telescope), radial velocities from the HARPS spectrograph on the ESO's 3.6\,m Telescope (prog ID 085.C-0393), and photometry from the robotic 60cm TRAPPIST telescope, all located at ESO, La Silla, Chile. The data is publicly available at the \textit{CDS} Strasbourg and on demand to the main author.}
\\ {\Large I -- Physical and orbital parameters, including spin--orbit angles, of two low-mass eclipsing binaries on opposite sides of the Brown Dwarf limit}
}
\author{Amaury H.M.J. Triaud\inst{1}
\and Leslie Hebb\inst{2}
\and David R. Anderson\inst{3}
\and Phill Cargile\inst{2}
\and Andrew Collier Cameron\inst{4}
\and Amanda P. Doyle\inst{3}
\and Francesca Faedi\inst{5}
\and Micha\"el Gillon\inst{6} 
\and Yilen Gomez Maqueo Chew\inst{5}
\and Coel Hellier\inst{3}
\and Emmanuel Jehin\inst{6} 
\and Pierre Maxted\inst{3}
\and Dominique Naef\inst{1}
\and Francesco Pepe\inst{1}
\and Don Pollacco\inst{5}
\and Didier Queloz\inst{1}
\and Damien S\'egransan \inst{1}
\and Barry Smalley\inst{3}
\and Keivan Stassun\inst{2}
\and St\'ephane Udry \inst{1}
\and Richard G. West\inst{7}
}

\offprints{Amaury.Triaud@unige.ch}

\institute{Observatoire Astronomique de l'Universit\'e de Gen\`eve, Chemin des Maillettes 51, CH-1290 Sauverny, Switzerland
\and Department of Physics and Astronomy, Vanderbilt University, Nashville, TN37235, USA
\and Astrophysics Group, Keele University, Staffordshire, ST55BG, UK
\and SUPA, School of Physics \& Astronomy, University of St Andrews, North Haugh, KY16 9SS, St Andrews, Fife, Scotland, UK
\and Astrophysics Research Centre, School of Mathematics \& Physics, QueenÕs University, University Road, Belfast, BT71NN, UK
\and Institut d'Astrophysique et de G\'eophysique, Universit\'e de Li\`ege, All\'ee du 6 Ao\^ut, 17, Bat. B5C, Li\`ege 1, Belgium
\and Department of Physics and Astronomy, University of Leicester, Leicester, LE17RH, UK
}

\date{Received date / accepted date}
\authorrunning{Triaud et al.}
\titlerunning{EBLM Project I - WASP-30b \& J1219--39b}

\abstract{This paper introduces a series of papers aiming to study the dozens of low mass eclipsing binaries (EBLM), with F, G, K primaries, that have been discovered in the course of the WASP survey. 
 Our objects are mostly single-line binaries whose eclipses have been detected by WASP and were initially followed up as potential planetary transit candidates. These have bright primaries, which facilitates spectroscopic observations
during transit  
and allows the study of the spin-orbit distribution of F, G, K+M eclipsing binaries through  the Rossiter--McLaughlin effect. 
\\Here we report on the spin-orbit angle of WASP-30b, a transiting brown dwarf, and improve its orbital parameters. We also present the mass, radius, spin-orbit angle and orbital parameters of a new eclipsing binary, J1219--39b (1SWAPJ121921.03--395125.6, TYC 7760-484-1), which, with a mass of $95\pm2\,M_\mathrm{jup}$, is close to the limit between brown dwarfs and stars. We find that both objects orbit in planes that appear aligned with their primaries' equatorial planes.  Neither primaries are synchronous. J1219--39b has a modestly eccentric orbit and is in agreement with the theoretical mass--radius relationship, whereas WASP-30b lies above it. 

 }

\maketitle

\section{Introduction}

The WASP consortium (Wide Angle Search for Planets) \citep{Pollacco:2006fj} has been operating from La Palma, Spain, and Sutherland, South Africa. Its main goal is to find transiting extrasolar planets. With more than 80 planets discovered, this is the most successful ground-based survey for finding short-period giant planets. Amongst the many planet candidates that WASP has produced are many `false positives', which here we regard as objects of interest, that have been shown by radial-velocity follow-up to be M dwarfs that eclipse F, G or K-dwarf companions. They are of a few Jovian radii in size and thus mimic a planetary transit signal very well. Because of the mass and low brightness of the secondaries,  they remain invisible, making them convenient objects for follow-up and study using the same photometry and radial-velocity techniques that are routinely used for exoplanets. Two A+M binaries have already been presented in \citet{Bentley:2009lr} and similar objects have been found by the OGLE survey \citep{Udalski:2007lr,Pont:2006lr} and by the HAT network \citep{Fernandez:2009fk}.

We have made a substantial effort to characterise these low-mass eclipsing binaries (the EBLM Project) in order to discover transiting brown dwarfs (such as WASP-30b \citep{Anderson:2011fk}) and also to complete the largely empty mass--radius diagram for stars with masses $< 0.4$ M$_\odot$. These objects explore the mass distribution separating stars from planets, or serve as extended samples to the exoplanets, especially with regards to their orbital parameters, long term variability and spin-orbit angles. Our results will be published in a series of papers, of which this is the first.\\

A primary goal of the EBLM Project is to address the M-dwarf radius  problem whereby current stellar evolution models underestimate the radii of M~dwarfs by at least 5\% and overestimate their temperatures by a few hundred degrees (e.g. \citep{Morales:2010fk, Morales:2009fj,Lopez-Morales:2007kx} and references therein). Thus we aim to substantially increase the number of M dwarfs with accurate masses, radii, and metallicities using a large sample of newly discovered eclipsing binaries comprising F, G, K primaries with M dwarf secondaries. The masses and radii results are inferred using F, G, K atmospheric and evolution models. Although model-dependent, the analysis of bright F, G, K + M dwarf eclipsing binaries promises large numbers of masses and radii of low-mass stars over the entire range of M dwarfs down to the hydrogen-burning limit. They will have accurate metallicity determination, and cover a wide range of activity levels. A combined analysis of the radial-velocity curve and light curve permits to deduce the masses and radii, while an accurate system metallicity can be determined from the F, G, K primary star. Furthermore, activity can be determined indirectly through knowledge of the rotation-activity relation \citep{Morales:2008qy} combined with $V\,\sin\,i_\star$ from measurements or by deduction when the systems are tidally synchronised.\\


\citet{Holt:1893fk}, in proposing a method to determine the rotation of stars prior to any knowledge about line broadening, predicted that when one star of a binary eclipsed the other it would first cover the advancing blue-shifted hemisphere and then the receding red-shifted part. This motion would create a colour anomaly perceived as a progressive red-shift of the primary's spectrum followed by a blue-shift, thus appearing as a symmetric radial-velocity anomaly on top of the main Doppler orbital motion of the eclipsed star's lines. This effect was first observed by \citet{Rossiter:1924qy} and \citet{McLaughlin:1924uq}, though with some evidence of its presence noted earlier by \citet{Schlesinger:1910fj} (p134). Holt's  idea was correct but only under the assumption that both stars orbit in each other's equatorial plane. In the case of a non-coplanar orbital motion the radial velocity effect is asymmetric (see e.g. \citet{Gimenez:2006kx} or \citet{Gaudi:2007vn})

Recent observations  of this effect in transiting extrasolar planets (e.g. \citet{Queloz:2000rt,Winn:2005ys,Hebrard:2008mz,Winn:2009lr,Triaud:2010fr,Moutou:2011cr,Brown:2012lr} and references therein) have shown that the so-called \textit{hot Jupiters}, gas giant planets on orbits $<$ 5 days, have orbital spins on a large variety of angles with respect to the stellar spin axis,  the most dramatic cases being on retrograde orbits. While it was previously thought that hot Jupiters had migrated from their formation location to their current orbits via an exchange of angular momentum with the protoplanetary disc, they are now thought to have been dynamically deflected onto highly eccentric orbits that then circularised via tidal friction. There are various ways in achieving this, such as planet--planet scattering \citep{Rasio:1996ly,Nagasawa:2008gf,Wu:2011ul} and Kozai resonances \citep{Kozai:1962qf,Wu:2007ve,Fabrycky:2007pd,Naoz:2011bh}. These could be triggered by environmental effects in their original birth clusters such as fly-bys \citep{Malmberg:2007lq,Malmberg:2011dq}, by an additional, late, inhomogenous mass collapses in young systems \citep{Thies:2011yq}, or during the planet formation process itself \citep{Matsumura:2010ve,Matsumura:2010ul}. 

Several patterns have emerged in the planetary spin--orbit angle data, including: a lack of aligned systems whose host stars have $T_\mathrm{eff} > 6250$ K \citep{Winn:2010rr}; a lack of inclined systems older than 2.5--3 Gyr \citep{Triaud:2011fk}; and a lack of retrograde system for secondaries $> 5$ M$_\mathrm{Jup}$ \citep{Hebrard:2011fk,Moutou:2011cr}. To help confirm this latter trend, one could measure the Rossiter--McLaughlin effect in several heavy planets, but those are rare. It is thus easier to extend the mass range to low-mass stars, hoping to further our understanding of the planetary spin--orbit angle distribution. \\

The fact that hot Jupiters can be on inclined orbits raises the question about the inclinations of close binary stars. As proposed by \citet{Mazeh:1979eu},  close binaries, especially those with large mass differences, might form via the same dynamical processes that have been proposed for hot Jupiters, i.e., gravitational scattering followed by tidal friction. In fact, \citet{Fabrycky:2007pd} primarily address the formation of close binaries; the possible application to exoplanets comes later. That paper was motivated by observational results, notably presented by \citet{Tokovinin:2006la}, showing that at least 96\,\% of close binaries are accompanied by a tertiary component, supporting the appeal to the Kozai mechanism as described in \citet{Mazeh:1979eu}. It has been argued that objects as small as 5 M$_\mathrm{jup}$ could be formed as stars do \citep{Caballero:2007mz}, while objects as massive as 20 or 30~M$_\mathrm{jup}$ could be created by core collapse, in the fashion expected for planets \citep{Mordasini:2009gf}. Rossiter--McLaughlin observations bridging the mass gap between planets and stars could eventually help in separating or confirming both proposed scenarii.\\

Even though attempts have been made to model the Rossiter--McLaughlin effect (e.g. \citet{Kopal:1942ai,Hosokawa:1953kl}) no systematic, quantified and unbiased survey of the projected spin--orbit angle $\beta$ in binary  star systems can be found in the literature. Only isolated observations of nearly aligned systems have been reported. \citet{Kopal:1942qe} mentions a possibly asymmetric Rossiter--McLaughlin effect (or rotation effect as it was then known) leading to an estimated misaligned angle of $15^{\circ}$ observed in 1923 in the Algol system, but that was presented as aligned by \citet{McLaughlin:1924uq}. \citet{Struve:1950tg} (p 125) writes that the rotation effect had been observed in a 100 systems without citing anyone. \citet{Slettebak:1985dp} is a good source of citations about this epoch.  \citet{Worek:1996hc} and \citet{Hube:1982sp} are two examples of more recent observations of the Rossiter--McLaughlin effect. The rotation effect was also used for cataclysmic variables to determine if the accreting material comes from a disc in a  plane similar to the binary's orbital plane \citep{Young:1980ij}. 

It has to be noted that, early on, the Rossiter--McLaughlin effect was used as a tool to measure the true rotation of a star, hence creating a bias against reporting misaligned systems. Furthermore the precision and accuracy of instrumentation, data extraction and analysing technique of that time prevented the observation of the Rossiter--McLaughlin effect for slowly rotating stars, further biasing detections of the effect towards synchronously rotating binaries, which could have tidally evolved to become aligned \citep{Hut:1981kx}. 

In addition, the capacity to accurately model blended absorption lines of double-lined binaries (SB2) during transit has only been developed recently. Thus, most people that studied binaries chose not to observe during eclipses. Modelling eclipsing SB2 has recently been developed in \citet{Albrecht:2007th}, and used by \citet{Albrecht:2009fy} for the case of DI Herculis, explaining its previously abnormal apsidal motion: both stars orbit above each other's poles. These measurements are being followed by a systematic and quantified survey of spin--orbit measurements for SB2s of hot stars with similar masses (the BANANA project, \citet{Albrecht:2011bs}). Another contemporary result is presented in an asteroseismologic paper by \citet{Desmet:2010fv}.   

We circumvent the  blended-line problem altogether by only observing the WASP candidates that turned out to be single-line binaries (SB1) while searching for extrasolar planets. Low-mass M dwarfs and brown dwarfs have a size similar to gas giants and appear to a first approximation like a planet: a dark spot moving over the disc of their primary. Thus, the low-mass eclipsing binaries found by transiting planet surveys  provide a  good sample to extend the work carried out on planets and provide a largely unbiased, quantified survey of spin--orbit angles for F, G or K~+~M binaries, complementary to the BANANA project.  The differences between our primaries will also allow us to probe the way tides propagate in convective or radiative stars \citep{Zahn:1977yq}. In stellar parameters and data treatment, our systems resemble the aligned pair Kepler-16 A\,\&\,B \citep{Doyle:2011vn,Winn:2011wd}, but with shorter periods. 

In this we first present our observations of WASP-30 and J1219--39 (1SWASPJ121921.03--395125.6, TYC 7760-484-1), then describe our models and how they were adjusted to fit the data, and how the error bars were estimated. We will then move to a discussion of the results.

\section{Observations}\label{sec:obs}

The discovery of WASP-30 was announced in \cite{Anderson:2011fk}. This is a transiting -- or eclipsing -- 61\,M$_\mathrm{Jup}$ brown dwarf on a 4.16-day orbit.  In our analysis we have used the data published in \cite{Anderson:2011fk} as well as new observations. The full sample comprises photometric observations from three facilities: 
the WASP-South photometry (four datasets totalling 17\,528 independent measurements) and the Gunn $r'$ \textit{Euler} photometry (one set of 250 points) were presented in \cite{Anderson:2011fk}. In addition we present 571 new photometric observations obtained in the $I+z$ band using the TRAPPIST telescope, covering the transit of 2010 October 15. We also gathered radial-velocity data: 32 spectra were observed using CORALIE (mounted on the Swiss  1.2m {\it Euler} Telescope) of which 16 have been published by \cite{Anderson:2011fk}. We also acquired 37 measurements using HARPS-South on the ESO 3.6m. 8 CORALIE and 16 HARPS measurements were obtained during the transits of 2010 October 15 and 2010 September 20, thus recording the Rossiter--McLaughlin effect.

%

J1219--39 is located at $\alpha = 12^\mathrm{h}\,19'\,21.03"$ and $\delta= -39^{\circ}\,51'\,25.6"$. Its name is a short version of its WASP catalogue entry. WASP-South observed a total of 22\,032 points in four series of photometric measurements obtained between 2006 May 04 and 2008 May 28. The automated \textit{Hunter} algorithm  \citep{Collier-Cameron:2007pb} found a periodic signal with period 6.76 days. This period was confirmed with 20 out-of-transit radial-velocity measurements obtained with CORALIE between 2008 August 03 and 2011 April 17. We also acquired an additional 54 measurements by observing nights during which three primary eclipses occurred (on 2010 May 13, 2010 July 13 and 2011 April 16). Several spectra were affected by bad weather conditions. Points with error bars above 20 m\,s$^{-1}$ were thus removed, leaving 61 RV points with an average error of 9.9\,m\,s$^{-1}$. Of these, 19 were taken during the Rossiter--McLaughlin effect. Because the aim of this paper is not about characterising the radius of this object but about orbital parameters, the WASP photometry is the only photometry we will use, which is sufficient to help in constraining the Rossiter--McLaughlin effect. This however does not prevent us from using the fact that the object eclipses to help get its mass and infer an estimate of its radius.

Additional details are located in the observational journal in the appendices, and a summary is displayed in table \ref{tab:obj}.

\section{Data Treatment}\label{sec:treat}




\begin{table}
\caption{Stellar parameters and abundances of our two primaries from spectral line analysis.}\label{tab:obj}
\begin{tabular}{lcc}
\hline
\hline
 & WASP-30A & J1219--39A \\
\hline
1SWASP&J235338.03--100705.1& J121921.03--395125.6\\
 2MASS&J23533805--1007049&J12192104--3951256\\
 TYC & 5834-95-1&7760-484-1\\
\hline
	& F8V & K0V\\
 $\alpha$ & $23^\mathrm{h}\,53'\,38.05"$&$12^\mathrm{h}\,19'\,21.03"$\\
 $\delta$   & $-10^{\circ}\,07'\,05.1" $&$-39^{\circ}\,51'\,25.6" $\\
 V mag	&		11.9		& 10.3	\\
 \\
$ T_\textrm{eff}$ (K)           & $6190\pm50$       & $5400\pm80$\\
$\log\,g$                               & $4.18\pm0.08$        & $4.5 \pm 0.1$ \\
\mictrb   \,(km\,s$^{-1}$)	& $1.1\pm0.1$		& $1.0\pm0.1$\\
\mactrb   \,(km\,s$^{-1}$)	 &$3.8\pm0.3$		&$1.4\pm0.3$	\\
$v\,\sin\,i_\star$   (km\,s$^{-1}$)& $12.1\pm0.5$         & $4.5\pm0.4$ \\
\\
{[Fe/H]}                & $0.07\pm0.08$ & $-0.23\pm 0.08$\\
{[Na/H]}   &   0.09 $\pm$ 0.04 & $-0.15 \pm0.05$\\ 
{[Ca/H]}   &   0.11 $\pm$ 0.12 &$-0.07 \pm 0.11$\\
{[Ti/H]}   &   0.02 $\pm$ 0.10 &$-0.08 \pm0.06$\\
{[Cr/H]}   &   0.11 $\pm$ 0.10 &$-0.17 \pm 0.0$\\
{[Ni/H]}   &   0.04 $\pm$ 0.08 &$-0.19 \pm 0.08$\\
{[Mg/H]}&-&$-0.10\pm 0.05$\\
{[Al/H]} & - & $-0.17 \pm 0.08$\\
{[Si/H]} &-& $-0.13 \pm 0.07$\\
{[Sc/H]} &-&$ -0.18 \pm 0.06$\\
{[V/H]} &-&$ -0.01 \pm 0.09$\\
{[Mn/H]} &-&$ -0.11 \pm 0.13$\\
{[Co/H]} &-&$-0.11 \pm 0.06$\\
$\log A$(Li)  &   2.98 $\pm$ 0.04 & $<$ 0.15\\
\\
\multicolumn{3}{l}{\textit{from the \citet{Torres:2010uq} relation}}\\
$M_\star$ ($M_\odot$)             &  $1.28\pm0.09$    & $0.89\pm0.06$\\
$R_\star$ ($R_\odot$)             &  $1.51\pm0.17$    & $0.87\pm0.11$\\
\\
\multicolumn{3}{l}{\textit{facilities used \& number of observations used in the analysis}}\\
WASP-South [V+R]	& 17\,528 	& 22\,032\\
TRAPPIST [I+$z$]	& 571	& -\\
EulerCam	 [$r$']	& 250	& -\\
CORALIE			& 32	& 61\\
HARPS			& 37	&-\\
\\

\hline
\end{tabular}
\newline {\bf Note:}  Spectral Type estimated from \teff\
using the table in \citet{Gray:2008fj}.
\end{table}

\subsection{the WASP-South photometry}\label{sec:wasp}

We used standard aperture photometry as described in
Section 4.3 of \citet{Pollacco:2006fj} where  
a 3.5-pixel aperture around the source is used 
(with the source position taken from a catalogue). Sky subtraction comes from an annulus (with radii of
13 to 17 pixels). Regions around catalogued stars and cosmic rays are removed from that calculation. The pixel scale is 13.7" per pixel. To maximise photons, we observed in white light, only with a cut-off filter in the far red in order to reduce effects from fringing. This is a large bandpass approximating to V+R. \\

\begin{figure}
\centering
\includegraphics[width=9cm]{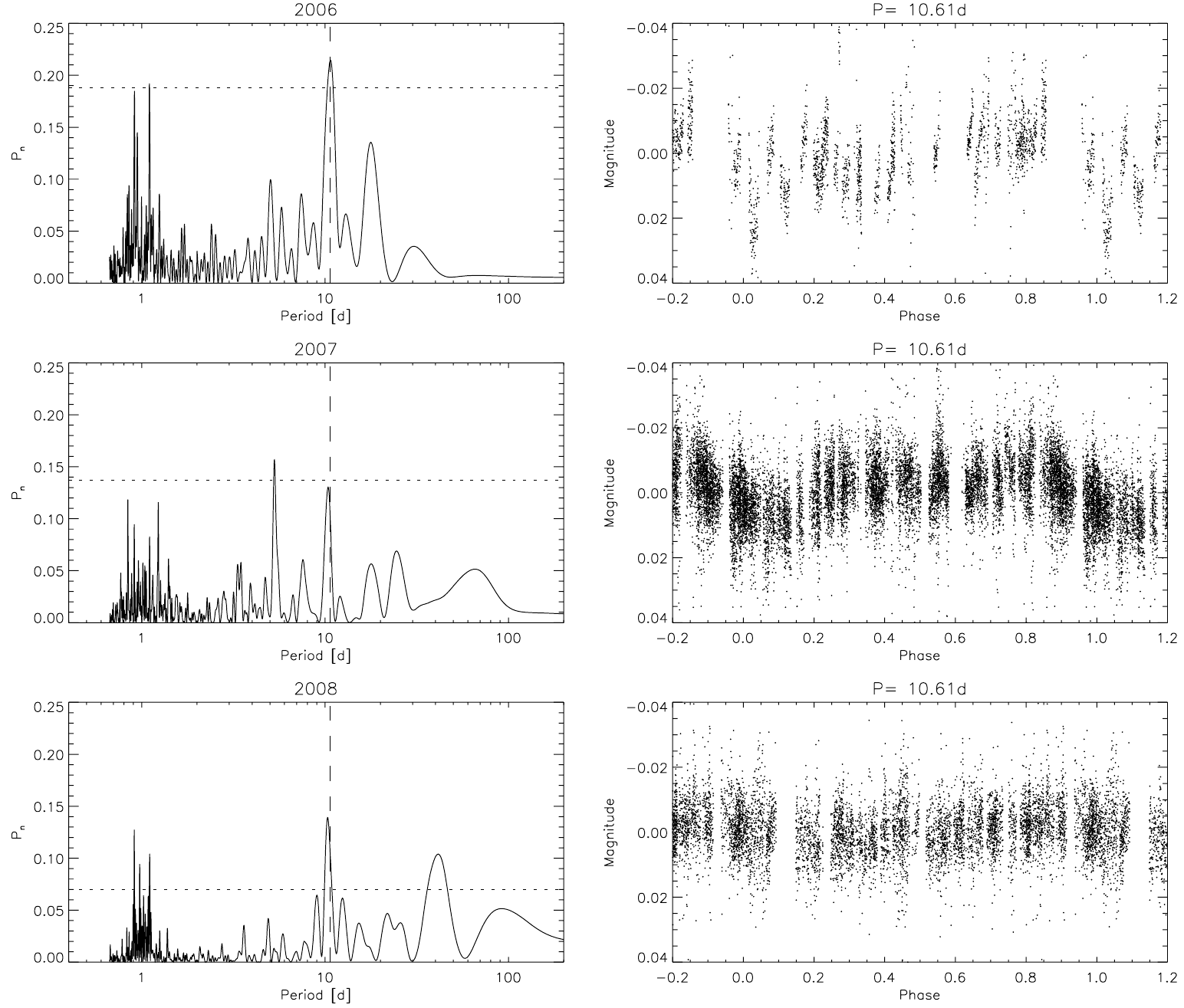}
\caption{{\it Left panels} Periodograms for the WASP data from the different
observing seasons for J1219--39. Horizontal lines indicate false alarm
probability levels FAP=0.01 and vertical lines show our assumed rotational
modulation period of P=10.6\,d. The year of observation in given in the title.
{\it Right panels} Lightcurves folded on the period P=10.61\,d.}
\label{fig:swlomb}
\end{figure}

We used the sine-wave fitting method described in \citet{Maxted:2011lr} to
search for any periodicity in the WASP lightcurves owing to the rotation of the
stars and caused by magnetic activity, i.e., star spots. Spot-induced variability is not expected to be coherent on long timescales as a consequence of
the finite lifetime of star-spots and of differential rotation in the photosphere 
so we analysed each season of WASP data separately.  We first subtracted a
simple transit model from the lightcurve. We then calculated periodograms over
4096 uniformly spaced frequencies from 0 to 1.5 cycles/day  (Fig. \ref{fig:swlomb}). The false-alarm
probability levels shown in these figures are calculated using a boot-strap
Monte Carlo method also described in \citet{Maxted:2011lr}. 

 For WASP-30 we analysed WASP lightcurves from 3 different seasons with
several thousand observations over about 100 days. These lightcurves show  no significant periodic out-of-transit variability. We examined the
distribution of amplitudes for the most significant frequency in each Monte
Carlo trial and used  these results to estimate a 95\,\% upper confidence
limit of 0.8\,milli-magnitude for the amplitude of any
periodic signal in these lightcurves.

The results  for our periodogram analysis of J1219--39 are shown in
Table~\ref{ProtTable}. Our interpretation is that we do
detect rotational modulation of J1219$-$39 and that in the 2007 data the pattern of spots results in the strongest
signal being seen at $P_{\rm rot}/2$, where $P_{\rm rot}$ is the
rotation period of the star at the latitude of the star spots. This leads to $P_{\rm
rot}=10.61\pm0.05$\,d  calculated from the unweighted mean and standard error on the mean from the three seasons. The periodograms for these data and the lightcurves
folded on this value of  $P_{\rm rot}$  are shown in Fig.~\ref{fig:swlomb}. 

\begin{table} 
\caption{Frequency analysis for J1219--39. $N$
is the number of observations, $P$ is the period corresponding to the
strongest peak in the periodogram, Amp is the amplitude of the best-fit sine
wave in milli-magnitudes and FAP is the false-alarm probability. } 
\label{ProtTable} 
\begin{tabular}{lrrrr} 
\hline 
\noalign{\smallskip}
\multicolumn{1}{l}{Year} & 
\multicolumn{1}{l}{N} & 
\multicolumn{1}{l}{P (day)} & 
\multicolumn{1}{l}{Amp (mmag)} & 
\multicolumn{1}{l}{FAP}\\ 
\noalign{\smallskip}
\hline
\noalign{\smallskip}
2006       & 1855 &  10.710 & 6 & 0.027 \\
2007       & 12104&   5.292 & 5 & 0.004 \\
2008       & 6019  &10.300	&3	&0.001\\
\noalign{\smallskip}
\hline 
\end{tabular} 
\end{table}

\subsection{the TRAPPIST $I+z$-band photometry}\label{subsec:trap}

\begin{figure}
\centering
\includegraphics[width=8.9cm]{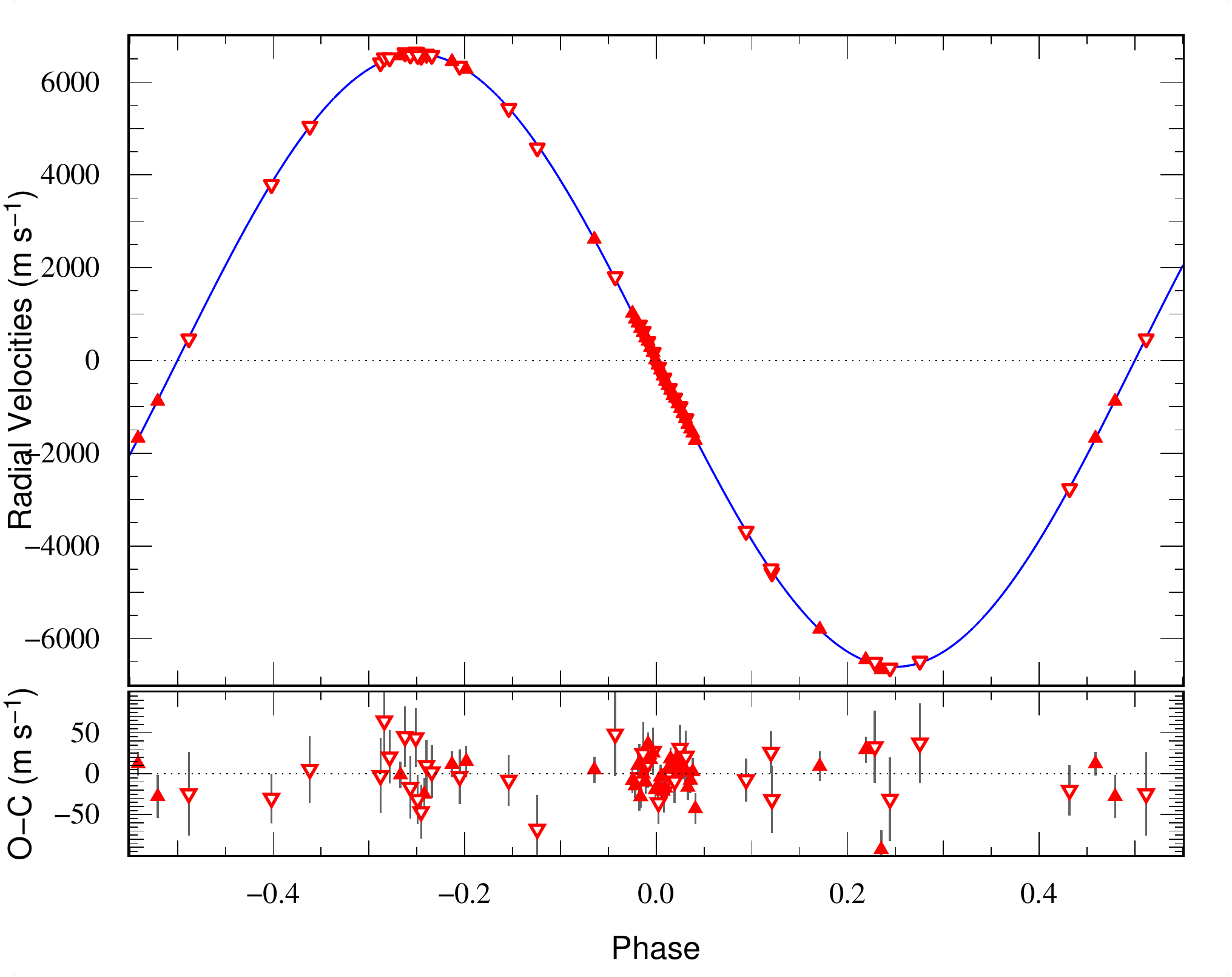}
\includegraphics[width=9cm]{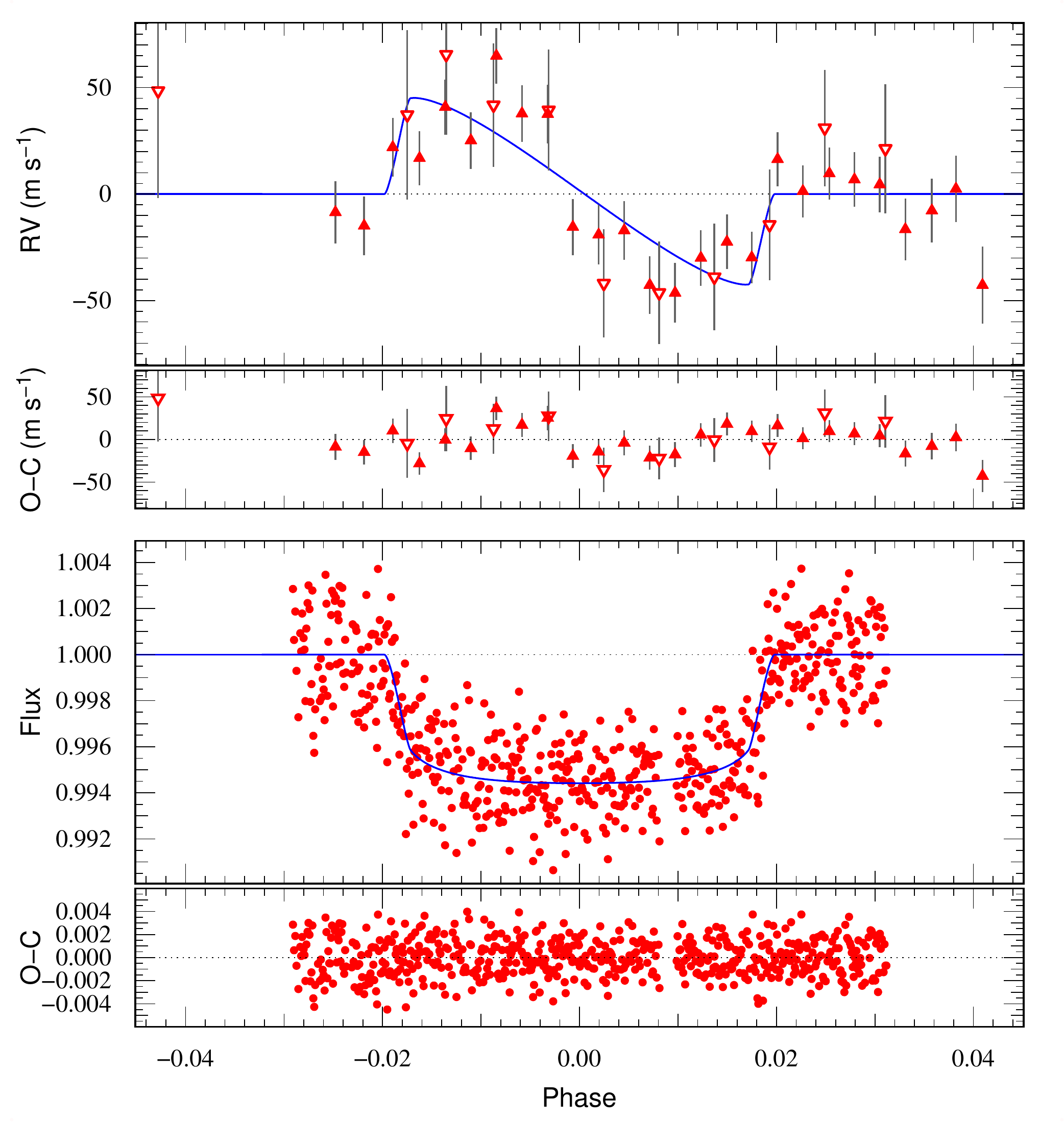}
\caption{top: Radial velocities on WASP-30 plotted with a circular Keplerian model and their residuals. CORALIE data is shown as inverted, empty, triangles. HARPS is show as upright triangles. Middle: zoom on the Rossiter--McLaughlin effect. Bottom: TRAPPIST $I+z$ photometry and model over-plotted. The interruption of the observations is due to a telescope meridian flip.}
\label{fig:wasp30}
\end{figure}

A complete transit  of WASP-30 was observed  with the robotic 60cm telescope TRAPPIST\footnote{http://arachnos.astro.ulg.ac.be/Sci/Trappist}) \citep{Gillon:2011fu,Jehin:2011dk}. Located at La Silla ESO observatory (Chile), TRAPPIST  is equipped with a 2K $\times$ 2K Fairchild 3041 CCD camera that has a 22' $\times$ 22' field of view (pixel scale = 0.64"/pixel). The transit of WASP-30 was observed on the night of 2010 October 15. The sky conditions were clear. We used the 1x2 MHz read-out mode with 1 $\times$ 1 binning, resulting in a typical read-out + overhead time  and read noise of 8.2 s and 13.5 $e^{-}$, respectively. The integration time was 30s for the entire night. We observed through a special $I+z$ filter that has a transmittance of zero below 700nm, and $>$ 90\% from 750nm to beyond 1100nm. The telescope was  defocused to average pixel-to-pixel sensitivity variations and to optimise the duty cycle, resulting in a typical full width at half-maximum of the stellar images of $\sim$6 pixels ($\sim$3.8"). The positions of the stars on the chip were maintained to within a few pixels over the course of two timeseries, separated by a meridian flip, thanks to the `software guiding' system that regularly derives an astrometric solution from the most recently acquired image and sends pointing corrections to the mount if needed. After a standard pre-reduction (bias, dark, flat field), the stellar fluxes were extracted from the images using the {\tt IRAF/DAOPHOT}  aperture photometry software \citep{Stetson:1987kl}. Several sets of reduction parameters were tested, and we kept the one giving the most precise photometry for the stars of brightness similar to WASP-30. After a careful selection of reference stars, differential photometry was obtained \citep{Gillon:2012fj}. The data are shown in figure \ref{fig:wasp30}. Because of a meridian flip inside the transit the photometry was analysed as two independent timeseries.

\subsection{the radial velocity data}\label{subsec:rv}

\begin{figure}
\centering
\includegraphics[width=8.9cm]{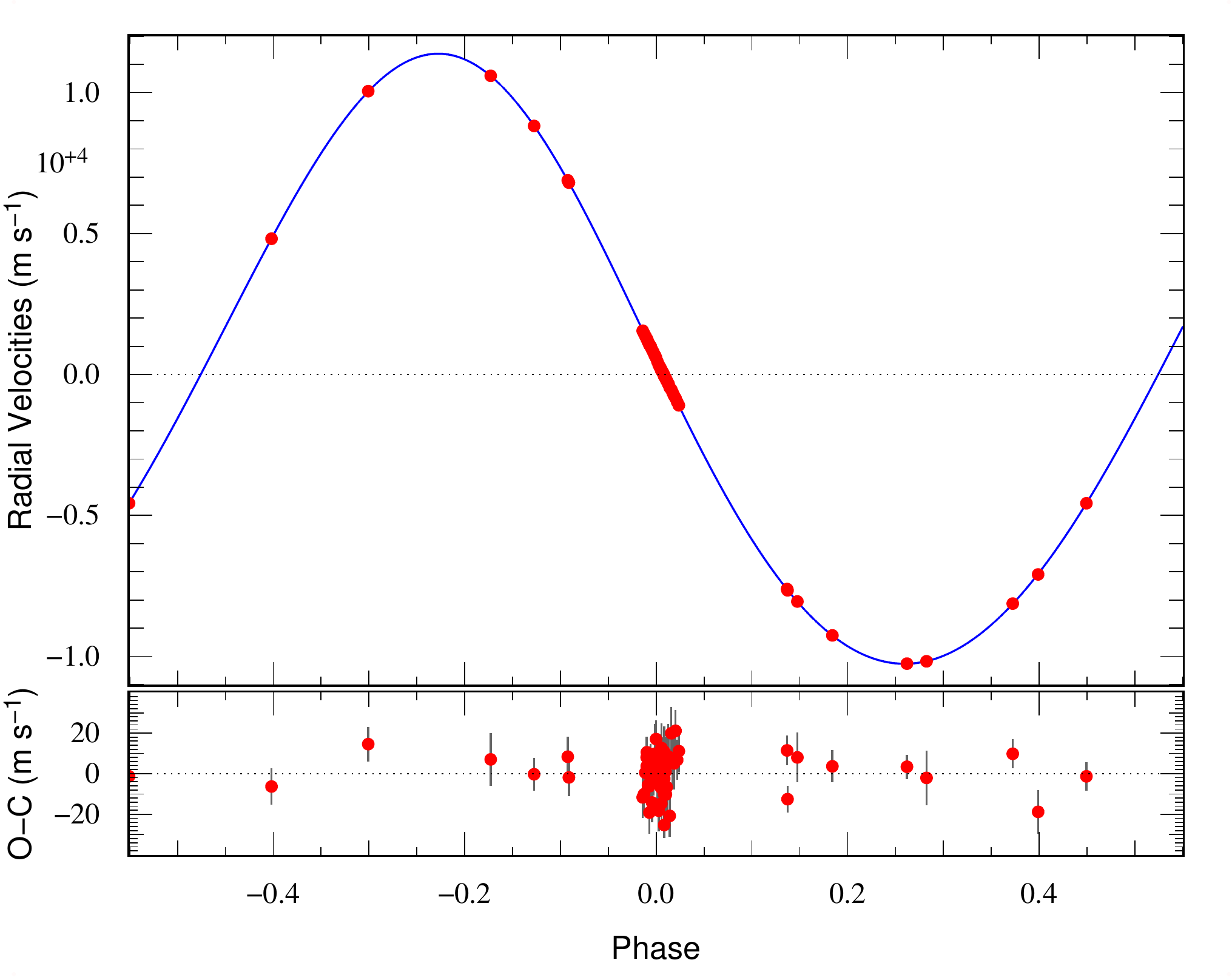}
\includegraphics[width=9 cm]{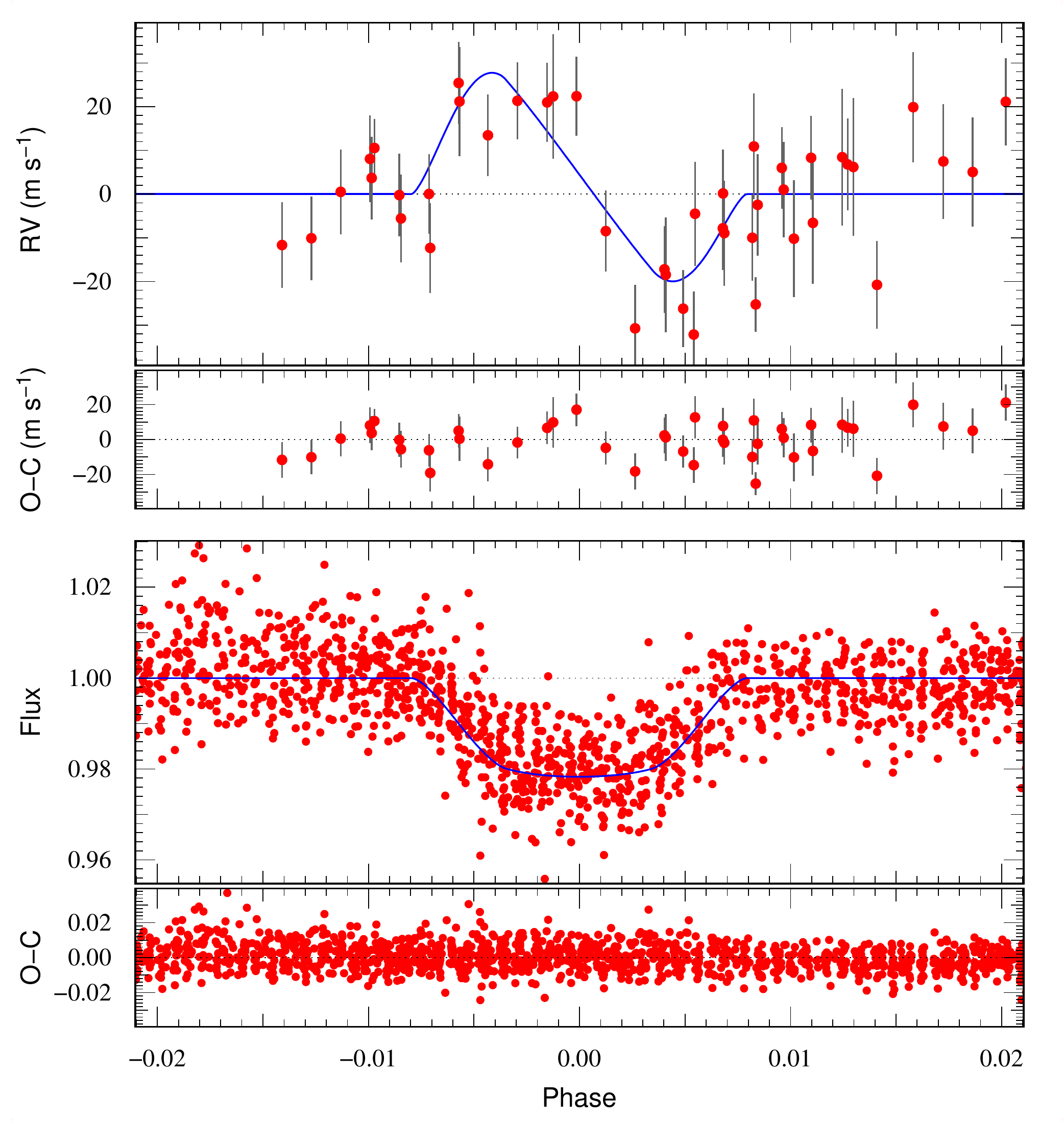}
\caption{top: CORALIE radial velocities on J1219--39 plotted with an eccentric Keplerian model and their residuals.  Middle: zoom on the Rossiter--McLaughlin effect. Bottom: WASP photometry and model over-plotted.}
\label{fig:SW1219}
\end{figure}

The spectroscopic data were reduced using the online Data Reduction Software (DRS) for the HARPS instrument. The radial-velocity information was obtained by removing the instrumental blaze function and cross-correlating each spectrum with a mask. This correlation was compared with the Th--Ar spectrum used as a wavelength-calibration reference (see \citet{Baranne:1996qa} \& \citet{Pepe:2002lh} for details). The DRS has been shown to achieve remarkable precision \citep{Mayor:2009rw} thanks to a revision of the reference lines for thorium and argon by \citet{Lovis:2007ec}. A similar software package was used to prepare the CORALIE data. A resolving power $R=110\,000$ for HARPS provided a cross-correlation function (CCF) binned in $0.25$\,km\,s$^{-1}$ increments, while for the CORALIE data, with a lower resolution of 50\,000, we used $0.5$\,km\,s$^{-1}$. 
The CCF window was adapted to be three times the size of the full width at half maximum (FWHM) of the CCF. 

$1\,\sigma$ error bars on individual data points were estimated from photon noise alone. HARPS is stable in the long term to within 1\,m\,s$^{-1}$ and CORALIE to better than 5\,m\,s$^{-1}$. These are smaller than our individual error bars, and thus were not taken into account.

As in the initial discovery paper, for WASP-30, a G2 mask was used.
In the case of  J1219--39 we used a K5 mask to extract the radial-velocity information.

Several points were removed from the analysis. For WASP-30, we excluded a mistakenly obtained series of 13 short CORALIE spectra taken during bad weather; the error bars are all above $> 100$ m\,s$^{-1}$ when no other data has errors  $> 50$ m\,s$^{-1}$. On SW1219--39b bad weather also affected observations, notably during one of the Rossiter--McLaughlin sequences. All measurements with error bars $> 20$ m\,s$^{-1}$ have been removed from the analysis as they show a clear jump in precision from other measurements. All the rejected radial velocities are nevertheless presented in the journal of observations, in the appendices, and are clearly indicated.

\section{Spectral analysis}\label{sec:spectral}

The analysis was performed following
the methods detailed in \citet{Gillon:2009uq} with standard pipeline
reduction products  used in the analysis. The \halpha\ line was used to
determine the effective temperature (\teff). The surface gravity (\logg) was
determined from the Ca~{\sc i} lines at 6162{\AA} and 6439{\AA}
\citep{Bruntt:2010qy}, along with the Na~{\sc i} D lines. Additional \teff\ and \logg\ diagnostics were performed using the Fe lines. An ionisation balance between Fe~{\sc i} and Fe~{\sc ii} was required, along with a null dependence of the abundance on either equivalent width or excitation potential \citep{Bruntt:2008fk}. The parameters obtained from the analysis are listed in Table~\ref{tab:obj}. The elemental abundances were determined from equivalent width measurements of several clean and unblended lines. A value for microturbulence (\mictrb) was determined from Fe~{\sc i} using the method of \citet{Magain:1984yq}. The quoted error estimates include that given by the uncertainties in \teff, \logg\ and \mictrb, as well as the scatter due to measurement and atomic data uncertainties. \\

A total of 37 individual HARPS spectra of WASP-30A were co-added to produce a
single spectrum with a typical S/N of around 120:1. Interstellar Na D lines are present in the spectra with an equivalent widths of $\sim$0.16\AA, indicating an
extinction of $E(B-V)$ = 0.05 using the calibration of \citet{Munari:1997zr}. The projected stellar rotation velocity (\vsini) was determined by fitting the
profiles of several unblended Fe~{\sc i} lines. A value for macroturbulence
(\mactrb) of 3.8 $\pm$ 0.3 km\,s$^{-1}$ was assumed, based on the calibration by
\citet{Bruntt:2010fk}. An instrumental FWHM of 0.07 $\pm$ 0.01~{\AA} was
determined from the telluric lines around 6300\AA. A best fitting value of
\vsini\ = 12.1 $\pm$ 0.5 ~km\,s$^{-1}$ was obtained. 

The lithium abundance would imply an age no more than $\sim$0.5~Gyr \citep{Sestito:2005ys} but stars with \teff\ similar to WASP-30A in M67 (5 Gyr old) have shown similar abundances  (Fig. 6 in \citet{Sestito:2005ys}). Those results are in agreement with the analysis of 16 co-added CORALIE spectra that was published by  \cite{Anderson:2011fk}. \\

A similar analysis was conducted on J1219--39A. Individual spectra were combined to a single spectrum of S/N typically 150:1. Using a macroturbulence \mactrb = $1.4\pm0.3$ km\,s$^{-1}$ \citep{Bruntt:2010fk} we obtain a \vsini\ = 4.5 $\pm$ 0.4 ~km\,s$^{-1}$. Using the macroturbulence calibration from \citet{Valenti:2005fj}, we find \mactrb = $3.4\pm0.3$ km\,s$^{-1}$, then we can infer \vsini\ = 3.3 $\pm$ 0.4 ~km\,s$^{-1}$. We will see later both those values are in contradiction with the $V\,\sin\,i_\star$, directly measured from the Rossiter--McLaughlin effect with the latter value being the closest. Here no Lithium can be detected for an equivalent width $< 1\mathrm{m}\AA$. The cores of the Ca H \& K lines show some emission indicative of stellar activity and in agreement with the detection of spot-induced variability (section \ref{sec:wasp}).\\

We will make a distinction in this paper between $v\,\sin\,i_\star$, the projected rotational velocity of the star computed by estimating the stellar line broadening, and the $V\,\sin\,i_\star$, which is the same physical quantity, but obtained directly from the amplitude of the Rossiter--McLaughlin effect. We also distinguish $i_\star$ the inclination of the stellar spin, from $i$, the inclination of the orbital spin of the companion.

\section{Model adjustment}\label{sec:fitting}

Simple Keplerian models were fitted to the radial-velocity data simultaneously with transit models from \citet{Mandel:2002kx} fitted to the photometry, and a Rossiter--McLaughlin model by \citet{Gimenez:2006kx} fitted to RV points falling within transit/eclipse. We used a quadratic limb-darkening law, and obtained parameters from \citet{Claret:2004fk} to apply to the photometry. For the Rossiter--McLaughlin effect, we applied parameters derived by Claret for HARPS \citep{Triaud:2009qy}. A Markov Chain Monte Carlo was used to compare the data and the models and explore parameter space to find the most likely model with robust confidence intervals on each parameter. Having only one set of parameters for all datasets  ensures the parameter distributions are consistent with all of the data. The algorithm is widely described in several planet-discovery papers from the WASP consortium (e.g. \citet{Collier-Cameron:2007fj,Gillon:2009qy,Anderson:2011fr,Triaud:2011qy}). The same method is used here with one important difference:\\

While adjusting for planets, many authors have used the property that the mass of the planet is much less than the mass of the star ($M_\mathrm{2} << M_\star$). This assumption is made in several places: in calculating the mass ratio from the mass function, the radii from the scaled radii $R_\mathrm{2}/a$ and $R_\star /a$, and in calculating the stellar density $\rho_\star$.  This last parameter is important since, being more precise than the traditional $\log g$, it is widely used to infer stellar parameters \citep{Sozzetti:2007fk}.

Three methods are often used in the literature involving Markov Chain Monte Carlo fitting algorithms that use the mean stellar density to obtain $M_\star$. One can fit the transit to obtain $\rho_\star$, use it to infer the stellar mass by interpolating of stellar tracks, and then insert the new $M_\star$  back into a chain, as employed by \citet{Hebb:2009lr}. Alternatively \citet{Enoch:2010lr} have devised an empirical relation based on the Torres relation \citep{Torres:2010uq} which also delivers $M_\star$. Finally one can mix both previous methods and estimate $M_\star$ at every step of an MCMC by using $\rho_\star$ to interpolate within theoretical stellar tracks as shown in \citet{Triaud:2011qy} and in \citet{Gillon:2012fj}. $\rho_\star$ is defined, from Kepler's law as:

\begin{equation}
{M_\star \over {R_\star^3}} = { 4 \pi^2 \over G P^2 } \left({a \over R_\star}\right)^3 -{M_\mathrm{2} \over R_\star^3}
\end{equation}

In our case, the second term can no longer  be considered null. In order to still be able to use the more precise $\rho_\star$ over $\log g$, we proceeded as follow: for every step in the MCMC we use the transit geometry to estimate the secondary's orbital inclination $i$. Then the mass function \citep{Hilditch:2001uq}

\begin{equation}
f(m) = (1 - e^2)^{3/2} \,{P\,K^3 \over 2 \pi \, G}
\end{equation}

is estimated. It can also be written as
\begin{equation}
f(m) = { (M_\mathrm{2} \,\sin i )^3 \over (M_\star + M_\mathrm{2} )^{2}}.
\end{equation}

Equating both, we can numerically solve for $M_\mathrm{2}$ assuming $M_\star$ (for example at the start of the chain, from the Torres relation). The orbital separation can then be estimated, and, having $R_\star /a$ from the transit signal, we obtain $R_\star$. We thus have gathered first estimates of all quantities necessary to  compute $\rho_\star$. This value is then combined with [Fe/H] and $T_\mathrm{eff}$ to give $M_\star$ from interpolating within stellar evolution tracks. $M_\star$ can then be used to re-estimate  $M_\mathrm{2}$ and $R_\star$, via the same path as outlined above, from which $R_\mathrm{2}$ is also determined, using the transit depth. This is repeated for each of the 2\,000\,000 steps of our MCMCs.\\

We use a new version of the Geneva stellar evolution tracks, described in \citet{Mowlavi:2012kx} and for the moment make the assumption that the primary is located on the main sequence. Because the steps falling outside the tracks are rejected, and provided the star is indeed on the main sequence, fitting using the tracks has the advantage of only allowing physically possible stars to be used in the MCMC. It has also the capacity to check and refine stellar parameters derived from spectral analysis, especially with respect to the lower boundary composed by the zero age main sequence.

Our Markov chains use the following jump parameters: $D$, the photometric transit/eclipse depth, $W$, its width, $b$, the impact parameter, $K$, the semi-amplitude of the radial velocity signal, $P$, the period, $T_0$, the transit mid-time point at the barycentre of the data (RV and photometric, weighted by their respective signal to noise). We also have two pairs of parameters: $\sqrt{e} \cos \omega$ \& $\sqrt{e} \cos \omega$ and $\sqrt{V \sin I} \sin \beta$ \& $\sqrt{V \sin I} \cos \beta$ where $e$ is the eccentricity, $\omega$ is the angle of periastron, $V \sin I$ is the rotation velocity of the star and $\beta$ is the projected spin-orbit angle. Those parameters are combined together to avoid inserting a bias in the determination of $e$ and $V \sin I$ (see \citet{Ford:2006yq,Triaud:2011vn} for details). [Fe/H] and T$_\mathrm{eff}$ are drawn randomly from a normal distribution taken from our spectral analysis. Normalisation factors for the photometry, and $\gamma$ velocities for each RV datasets, are not floating, but computed. For WASP-30, the RV data was cut into four datasets: the CORALIE data on the orbit, the CORALIE data during RM effect plus one measurement the night before and after, the HARPS data on the orbit, the HARPS RM effect plus one point the night before, and one the night after. Several chains are run to ensure, first, that convergence is achieved, but also to test the effects of different priors. 

From the jump parameters a number of physical parameters,  such as the masses and radii of both objects, can be computed. Useful, assumption-free parameters are also available, such as the secondary's surface gravity $\log\,g_2$, as noted in \citet{Southworth:2004uq}.

Results are taken as the modes of the posterior probability distributions. Errors for each parameter are obtained around the mode using the marginalised distribution and taking the 1-, 2- and 3-$\sigma$ confidence regions.

\section{Results}\label{sec:results}

\subsection{WASP-30b}

\begin{table}
\caption{Floating and computed parameters found for our two systems WASP-30A\&b and J1219--39A\&b. For clarity only the last two digits of the $1\,\sigma$ errors are shown.}\label{tab:params}
\begin{tabular}{lll}
\hline
\hline
Parameters $(units)$ & WASP-30 & J1219--39 \\
\hline
&\\
\textit{jump parameters} &\\
&\\
$P$ (days)                  	& $4.156739^{+(12)}_{-(10)}$  				&  6.7600098$^{+(34)}_{-(22)}$\\
$T_0$ (BJD-2\,450\,000) 	&$ 5443.06046 ^{+(43)}_{-(33)}$  	   		& 5187.72676$^{+(29)}_{-(41)}$    \\

$D$                              &$ 0.00494 ^{+(11)}_{-(13)}$    				&   0.02088$^{+(89)}_{-(69)}$\\
$W$ (days)                 &$ 0.1644 ^{+(13)}_{-(09) }$                   			& 0.1040$^{+(20)}_{-(20)}$   \\
$b$ (R$_{\star}$)       &$ 0.10 ^{+(0.12)}_{-(0.10)}$          		             	& 0.733$^{+(2)}_{-(31)}$     \\
$K$ (m\,s$^{-1}$)       &$ 6606.7 ^{+(4.7)}_{-(5.3)}$             		        	&10\,822.2$^{+(2.8)}_{-(3.1)}$    \\
$\sqrt{V\,\sin\,I}\,\cos\,\beta$	&	$ 3.40 ^{+(0.12)}_{-(0.24)}$		&1.61$^{+(0.13)}_{-(0.11)}$\\
$\sqrt{V\,\sin\,I}\,\sin\,\beta$	&	$ 0.5 ^{+(1.1)}_{-(1.6)}$			&0.13$^{+(0.13)}_{-(0.15)}$\\
$\sqrt{e}\,\cos\,\omega$		&			0 (fixed)				&0.21932$^{+(57)}_{-(48)}$\\
$\sqrt{e}\,\sin\,\omega$		&			0 (fixed)				&0.08537$^{+(85)}_{-(89)}$\\
      &        \\
      &        \\
\textit{derived parameters}       &      \\
&\\
$f(m)$ (M$_\odot)$			&  $0.00012418^{+(28)}_{-(29)} $		&0.000883709$^{+(71)}_{-(71)}$\\
$R_2 / R_{\star}$                         & $ 0.0704 ^{+(07)}_{-(10)} $ 			&0.1446$^{+(29)}_{-(25)}$ \\
 & \\
$R_{\star} / a$                               & $ 0.1164^{+(25) }_{-(12)} $		 	&0.0561$^{+(18)}_{-(23)}$ \\
$\rho_{\star}$   ($\rho_{\odot}$) & $ 0.466^{+(16)}_{-(29)} $        		&1.50$^{+(0.17)}_{-(0.17)}$ \\
$R_{\star}$  (R$_{\odot}$)          & $1.389^{(33)}_{(25)}$  				&0.811$^{+(38)}_{-(24)}$    \\
$M_{\star} $  (M$_{\odot}$)        & $ 1.249^{+(32)}_{-(36)}$       			&0.826$^{+(32)}_{-(29)}$ \\
 $\log\, g_\star$		(cgs)		& $ 4.250^{+(09)}_{-(18)}$			&4.523$^{+(39)}_{-(26)}$ \\
& \\
$R_2 / a$                               & $ 0.00821^{+(19) }_{-(18)} $			&0.00817$^{+(33)}_{-(48)}$  \\
$R_2$  (R$_\mathrm{Jup}$)       & $ 0.951^{+(28)}_{-(24)} $			&1.142$^{+(69)}_{-(49)}$   \\
$M_2$ (M$_\mathrm{Jup}$)                    & $62.5 ^{+(1.2)}_{-(1.2)}$		&95.4$^{+(1.9)}_{-(2.5)}$  \\
$\log\,g_2$	(cgs)			&	 $5.234 ^{+(19)}_{-(22)}$			&5.245$^{+(47)}_{-(42)}$ \\
&     \\
$a$ (AU)                        & $ 0.05534 ^{+(47)}_{-(51)} $     				&0.06798$^{+(83)}_{-(77)}$ \\
$i$   $(^{\circ})$                 & $ 89.43 ^{+(0.51)}_{-(0.93)} $        			&87.61$^{+(0.17)}_{-(0.18)}$   \\
$\beta$ ($^\circ$)		&	$7^{+(19)}_{-(27)} $					&4.1$^{+(4.8)}_{-(5.3)}$\\
$e$                                  & $  < 0.0044 $              						&0.05539$^{+(23)}_{-(22)}$  \\
$\omega$ ($^\circ$)		&		-							&21.26$^{+(0.21)}_{-(0.23)}$\\
$| \dot\gamma | $ (m\,s$^{-1}$\,yr$^{-1}$) &$<$ 53		&$<$ 10 \\
\\
$V\,\sin\,i_\star$ (km\,s$^{-1}$)&	$12.1^{+(0.4)}_{-(0.5)} $				&2.61$^{+(42)}_{-(35)}$\\
$T_\mathrm{eff}$ (K)			&	$6202^{+(42)}_{-(51)} $			&5412$^{+(81)}_{-(65)}$\\
{[Fe/H]}					&	$0.083^{+(69)}_{-(50)} $			&-0.209$^{+(70)}_{-(75)}$\\	
Age (Gyr)					&	$3.4^{+(0.3)}_{-(0.5)} $			& 6-12	\\
$\gamma_\mathrm{coralie} $ (km\,s$^{-1}$) & $7.9307^{+(22)}_{-(17)} $	&	$33.7971^{+(16)}_{-(15)} $	\\
$\gamma_\mathrm{harps}  $ (km\,s$^{-1}$) & $7.87472^{+(36)}_{-(31)} $	&-\\
\\
\hline

\end{tabular}

\end{table}

\begin{figure*}
\centering
\includegraphics[width=18cm]{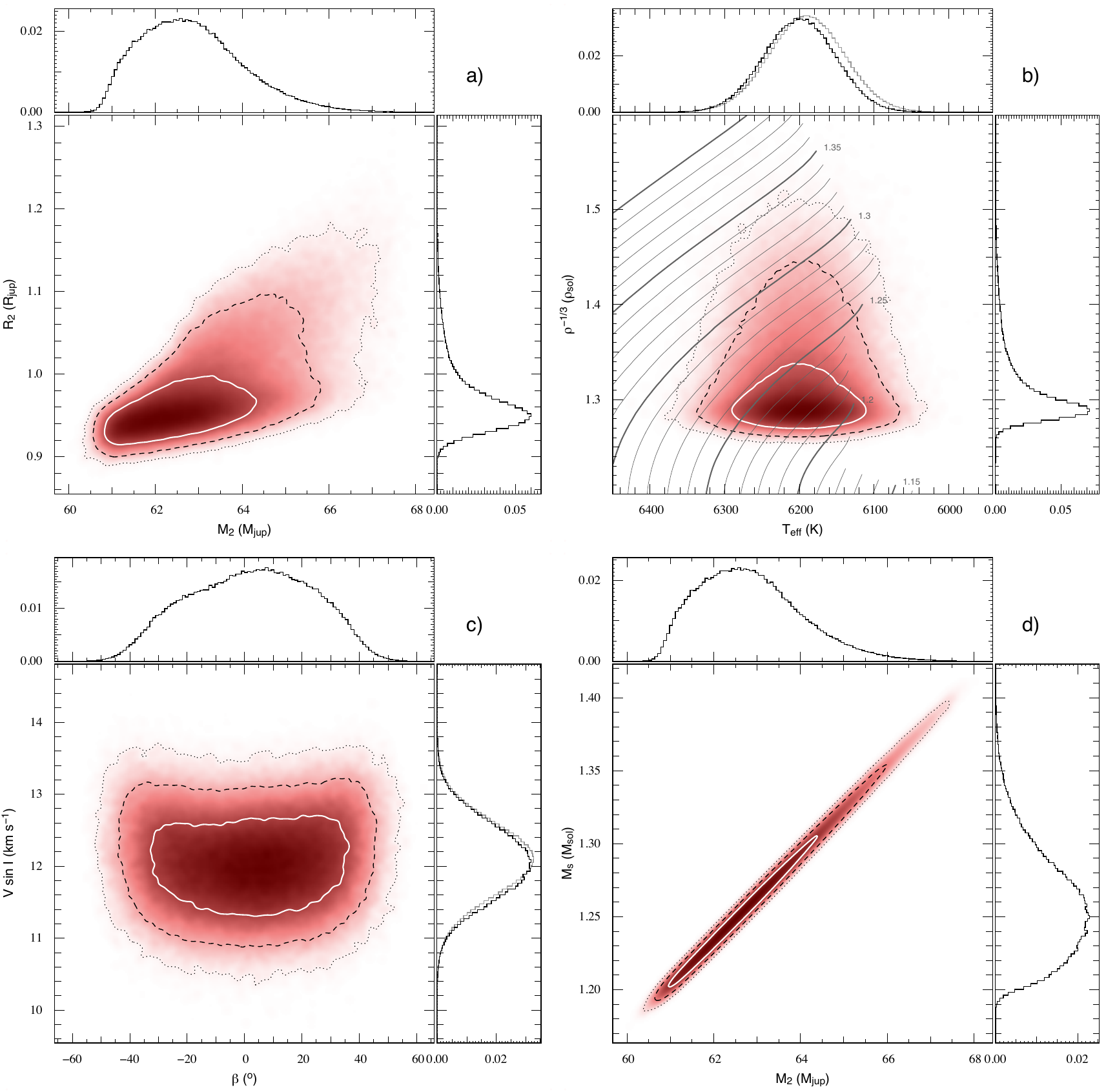}
\caption{WASP-30: The central panels show posterior probability-density distributions from the MCMC output, with contours at the 1-, 2- and 3-$\sigma$ confidence regions. The side panels show marginalised distributions as histograms in black. Where used, the priors are shown in grey. Panel (a) radius and mass of WASP-30b. b) modified Hertzsprung--Russell diagram over-plotted with the Geneva evolution tracks. Masses are indicated in $M_\odot$. c) $V\,\sin\,i_\star$ versus $\beta$ from fitting the Rossiter--McLaughlin effect. d)  dependence of the secondary's mass on our incomplete knowledge of the primary mass.}
\label{fig:w30param}
\end{figure*}

\begin{figure*}
\centering
\includegraphics[width=18cm]{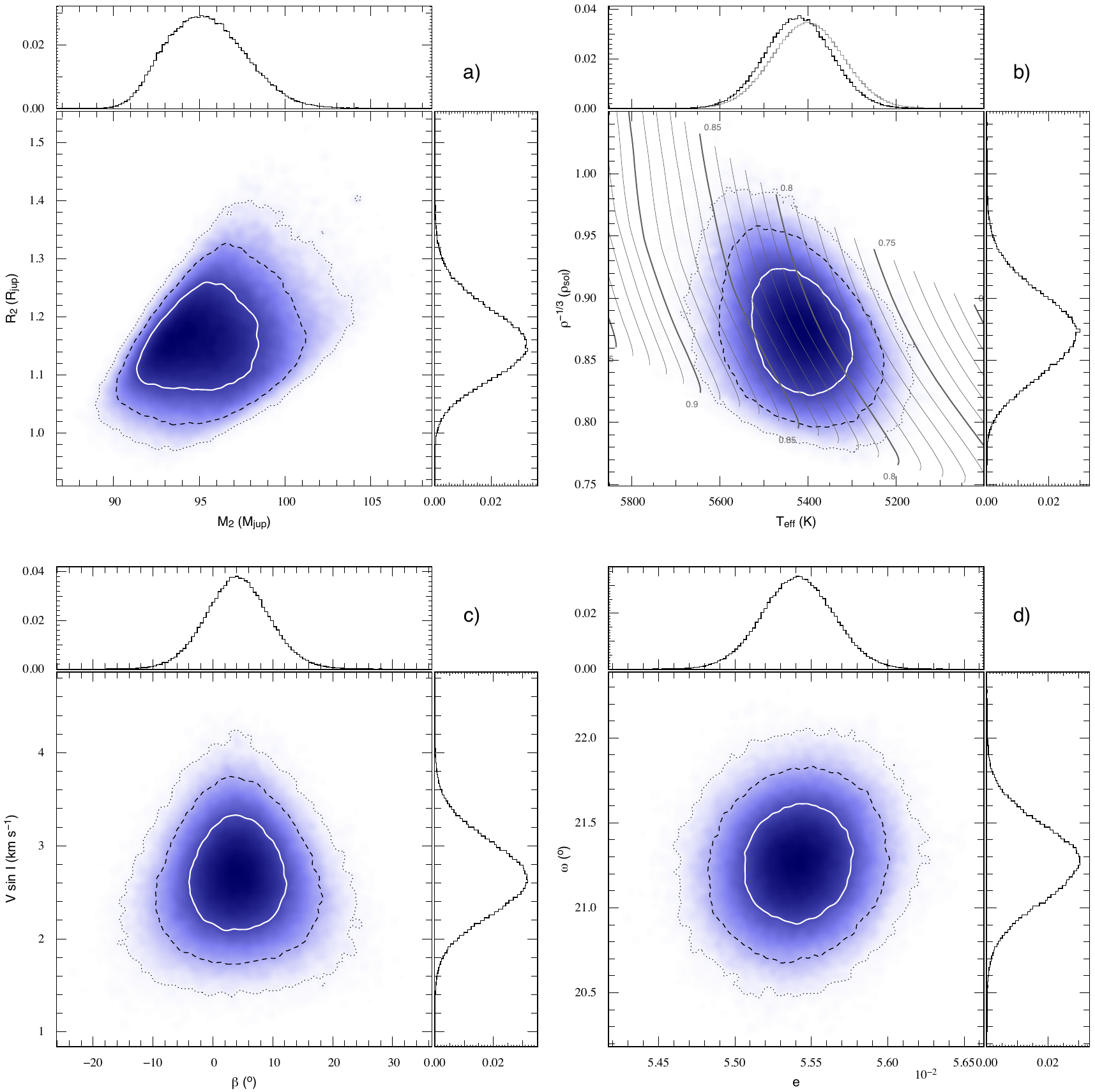}
\caption{J1219--39: The central panels show posterior probability-density distributions from the MCMC output, with contours at the 1-, 2- and 3-$\sigma$ confidence regions. The side panels show marginalised distributions as histograms in black. Where used, the priors are shown in grey. Panel (a) radius and mass of J1219--39b. b) modified Hertzsprung--Russell diagram  over-plotted with the Geneva evolution tracks. Masses are indicated in $M_\odot$. c) $V\,\sin\,i_\star$ versus $\beta$ from fitting the Rossiter--McLaughlin effect. d) the eccentricity $e$ and argument of periastron $\omega$.}
\label{fig:SW1219param}
\end{figure*}

Several chains were run on WASP-30b, exploring the effect that various priors could have on the end results. Overall the fit between the data and the models is good. In order to get a $\chi^2_\mathrm{reduced}$ close to 1, an extra contribution of 5 m\,s$^{-1}$ was added quadratically to the errors of the radial-velocity data. This stems mostly from one point in the HARPS data as well as from some high-cadence noise during the Rossiter--McLaughlin sequence. Nevertheless, we achieve a dispersion after the models are subtracted of 26.8 m\,s$^{-1}$ for a $\chi^2_\mathrm{reduced} = 1.47 \pm 0.21$.

The main difference between this analysis and that presented by \citet{Anderson:2011fk} is a slight change in mass and radius of WASP-30b,  arising from different choices for the estimation of the primary's parameters. \citet{Anderson:2011fk} used a Main-Sequence prior, which forced the photometric fit to be compatible with a smaller and less massive star. Upon relaxing the prior, those authors obtain a solution for the primary that is close to ours. The Main-Sequence prior is also the reason for a slightly different transit duration between the discovery paper and the current solution.  It forced a solution through the initial data which is no longer compatible with the addition of the TRAPPIST light curve. The data in the discovery paper only had WASP photometry and an Euler light curve that was imprecise during ingress. 
A small additional contribution comes from no longer making the planet approximation. 

Using priors on the values of $M_\star$ and $R_\star$ obtained using the Torres relation does not affect the result. Fits not using them are thus preferable as they constitute an independent measurement. WASP-30A is found to be a $1.25\pm0.03\,M_\odot$, $1.39\pm0.03\,R_\odot$ star at the end of its main-sequence lifetime. The values for $T_\mathrm{eff}$ and [Fe/H] from the output of the MCMC (table \ref{tab:params}) are entirely compatible with those presented in Table \ref{tab:obj}, meaning that we suffer little bias due to proximity of the star to the terminal-age main sequence. Thus, using these  stellar parameters we find that WASP-30b is a $62.5\pm1.2\,M_\mathrm{Jup}$ brown dwarf with a radius of $0.95\pm 0.03\,R_\mathrm{Jup}$. No eccentricity is detected. We can place a 95\,\%-confidence upper limit of  $e < 0.0044$. No additional acceleration is detected either. We calculated an upper constraint of $| \dot\gamma | < 53 $ m\,s$^{-1}$\,yr$^{-1}$.

The second feature of interest is the Rossiter--McLaughlin effect. Because of a low impact parameter the known degeneracy between $V \sin I$, $\beta$ and $b$  does not yield an unique solution (see for example \citet{Albrecht:2011rt} or \citet{Triaud:2011vn}). The application of a prior on $V \sin I$, using the value measured from spectral line broadening, prevents the MCMC from searching unphysical values of $V \sin I$. It also restricts the impact parameter $b$ from wandering too much, which otherwise was slightly affecting $\rho_\star$ and thus the primary's mass and radius.  We thus select the solutions using that prior. Full results are available in Table \ref{tab:params}.

WASP-30b is in a prograde orbit. Using the posterior distribution obtained for the stellar radius and the $V\,\sin\,i_\star$, the rotation period of the star is estimated at $5.9 \pm 0.3$ days, away from synchronisation. Any attempt to use higher values for  $V\,\sin\,i_\star$ would result in forcing the fit towards an inclined orbit.

\begin{figure}
\centering
\includegraphics[width=9cm]{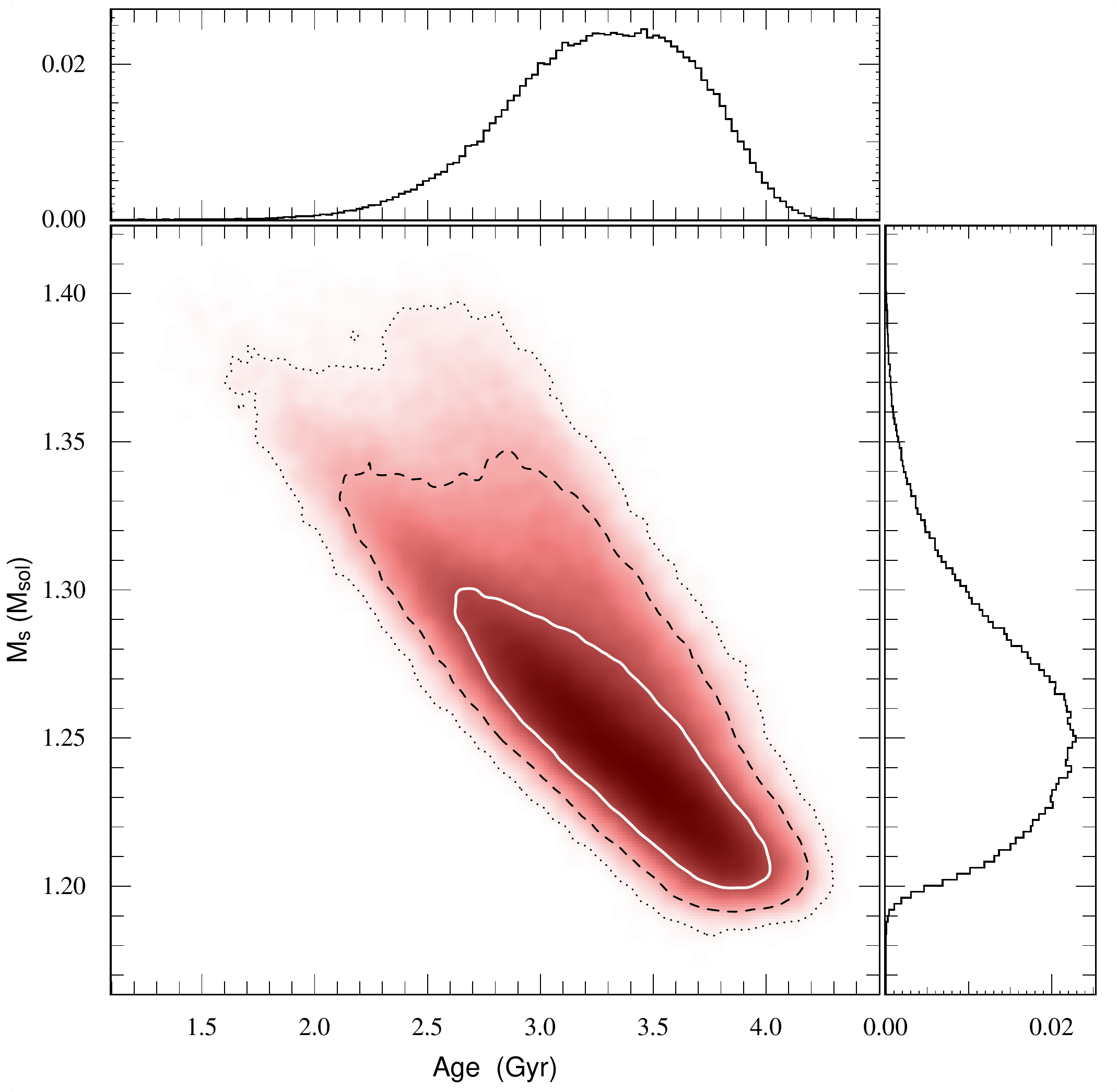}
\caption{The central panel shows the posterior probability-density distribution for stellar age against stellar mass for WASP-30. The contours show the 1-, 2- and 3-$\sigma$ confidence regions; the side panels show the marginalised distributions as histograms.  The data derive from interpolating using the mean stellar density, $T_\mathrm{eff}$ and metallicity, into the Geneva stellar evolution tracks \citep{Mowlavi:2012kx}.}
\label{fig:w30age}
\end{figure}

\subsection{J1219--39b}

The overall fit for J1219--39 is good: we obtain $\chi^2 = 82.4 \pm 12.8$ for 61 radial velocity points giving $\chi^2_\mathrm{reduced} = 1.53 \pm 0.23$ without adding a \textit{jitter} to the radial velocities. Errors on the photometry were adjusted to obtain a $\chi^2_\mathrm{reduced} =1$. 
The imposition of priors using the Torres relation on $M_\star$ and $R_\star$ does not affect the results. We thus adopt a prior-free chain. From the eclipse and spectroscopy, without any assumptions, we obtained $\log g_2 = 5.25 \pm 0.05$ indicating that an unresolved dense object is orbiting the primary. After careful analysis we find that J1219--39A is a $0.83\pm0.03\,M_\odot$, $0.81\pm0.03\,R_\odot$ star and its companion is a low mass star, of $0.091\pm0.002\,M_\odot$ and $0.117\pm0.006\,R_\odot$ ($95\pm2\,M_\mathrm{Jup}$ and $1.14\pm0.05\,R_\mathrm{Jup}$). The orbit is slightly eccentric ($e = 0.0554\pm0.0002$) while $\beta = 4^\circ\pm5$, showing good spin--orbit alignment. Full results are presented in Table \ref{tab:params}. Here too we do not detect any additional acceleration and can place an upper constraint with $| \dot\gamma | < 10 $ m\,s$^{-1}$\,yr$^{-1}$.

Claims of low eccentricities have been disputed in the past \citep{Lucy:1971uq}. Now, thanks to the high precision achieved with radial velocities, it is possible to measure extremely small orbital eccentricities. As a rule of thumb, one cannot conclusively detect eccentricity if the difference between the circular and eccentric model is smaller than the RMS of the residuals. This difference can be approximated by $2\,e\,K$. In our case we are well above that value. We nevertheless forced a circular model and find a much poorer fit reflected in a $\chi^2=46\,217\pm304$ instead of $\chi^2=62.2\pm11.1$ for the eccentric model (on the 42 points not affected by the RM effect). 

From fitting the Rossiter--McLaughlin effect we obtain an independent (prior-free)  distribution for stellar rotation peaking at $V\,\sin\,I = 2.6\pm0.4$ km\,s$^{-1}$. This value is significantly different from the value of $v\,\sin\,i_\star$ presented in Table \ref{tab:obj} and obtained from stellar line broadening. If we use $v\,\sin\,i_\star$ as a prior on $V\,\sin\,i_\star$, the fit of the Rossiter--McLaughlin effect worsens slightly, but stays within the natural noise variability.  
Under this prior, the most likely value becomes $V\,\sin\,I = 3.6\pm0.3$, in between the independent values. This matches the value of $v\,\sin\,i_\star$ obtained when using  a macroturbulence value from \citet{Valenti:2005fj}  instead of from \citet{Bruntt:2010fk}. We adopt the value of $V\,\sin\,I$ found from the Rossiter--McLaughlin effect alone, as it is a directly measured value, one that can be tested against macroturbulence laws.

Assuming coplanarity ($i = i_\star$) as indicated by $\beta$ and using the MCMC's posterior probabilities and the RM's $V\,\sin\,i_\star$, J1219--39A would have a rotation period of $15.2\pm2.1$ days. The solution using a prior on $V\,\sin\,i_\star$ gives a period of rotation of $11.7\pm1.0$ days, while when using instead the value of $v\,\sin\,i_\star$ in Table \ref{tab:obj} we obtain $9.0\pm0.9$ days. Since none of these is compatible with the orbital period, we have neither synchronisation between the secondary's orbital motion and the primary's rotation, nor pseudo-synchronisation. 

An analysis of the out-of-transit WASP data shows a recurring frequency at about 10.6 days on three seasons of data, presumably due to the rotation of stellar spots on the surface of the primary (see section \ref{sec:wasp}). This fourth possible rotation period is a direct observable. The discrepancy with the value obtained using the $V\,\sin\,i_\star$ is not understood, yet is interesting to note\footnote{this could lead to presume $ i_\star =42^\circ \pm 8$, indicating spin--orbit misalignment. The discrepancy between the different values of equatorial velocities prevents us from being sure.}. Further comparison between the $V\,\sin\,i_\star$ from the Rossiter--McLaughlin and photometric rotation periods seems warranted.




\section{Discussion \& Conclusions}

We announce the discovery of a new low-mass star whose mass and radius have been precisely measured and found to be at the junction between the stellar and substellar regimes. In addition, using observations with the CORALIE spectrograph, on the 1.2\,m Euler Telescope, and HARPS, on the ESO 3.6\,m, we have demonstrated the detection of the Rossiter--McLaughlin effect on two objects more massive than planets. These measurements are amongst the first to be realised on such objects. They will help study the dynamical events that could have led to the formation of binary systems where both components have a large mass difference, and may also provide a useful comparison sample to the spin--orbit angle distribution of hot Jupiters, as well as helping theoretical developments in the treatment of  tides, the main mechanism behind synchronisation, circularisation and realignment.

\citet{Hebrard:2011fk} and \citet{Moutou:2011cr} note that while hot Jupiters are usually found with a large variety of orbital angles, objects above 5 $M_\mathrm{jup}$ are not found on retrograde orbits. We extend the distribution of spin--orbit angle versus mass to beyond the planetary range. Both of our objects appear well aligned with their primary, further confirming that trend. Being around stars colder than 6250\,K, they also reinforce the pattern shown by \citet{Winn:2010rr} between orbital inclination and the primary's effective temperature. The age of WASP-30A and the alignment of WASP-30b also helps strengthen the pattern claimed in \citet{Triaud:2011fk} that systems older than 2.5--3 Gyr are predominantly aligned. It is interesting to note that while J1219--39b is on a slightly eccentric orbit, its orbital spin is aligned with its primary's rotational spin. This gives observational evidence that orbital realignment may be faster than orbital circularisation for these objects, in opposition with planets. Final tidal equilibrium has not been reached for either system as WASP-30A is not synchronised and J1219--39b is not circularised (nor synchronised) \citep{Hut:1980bh}.\\

\begin{figure}
\centering
\includegraphics[width=9cm]{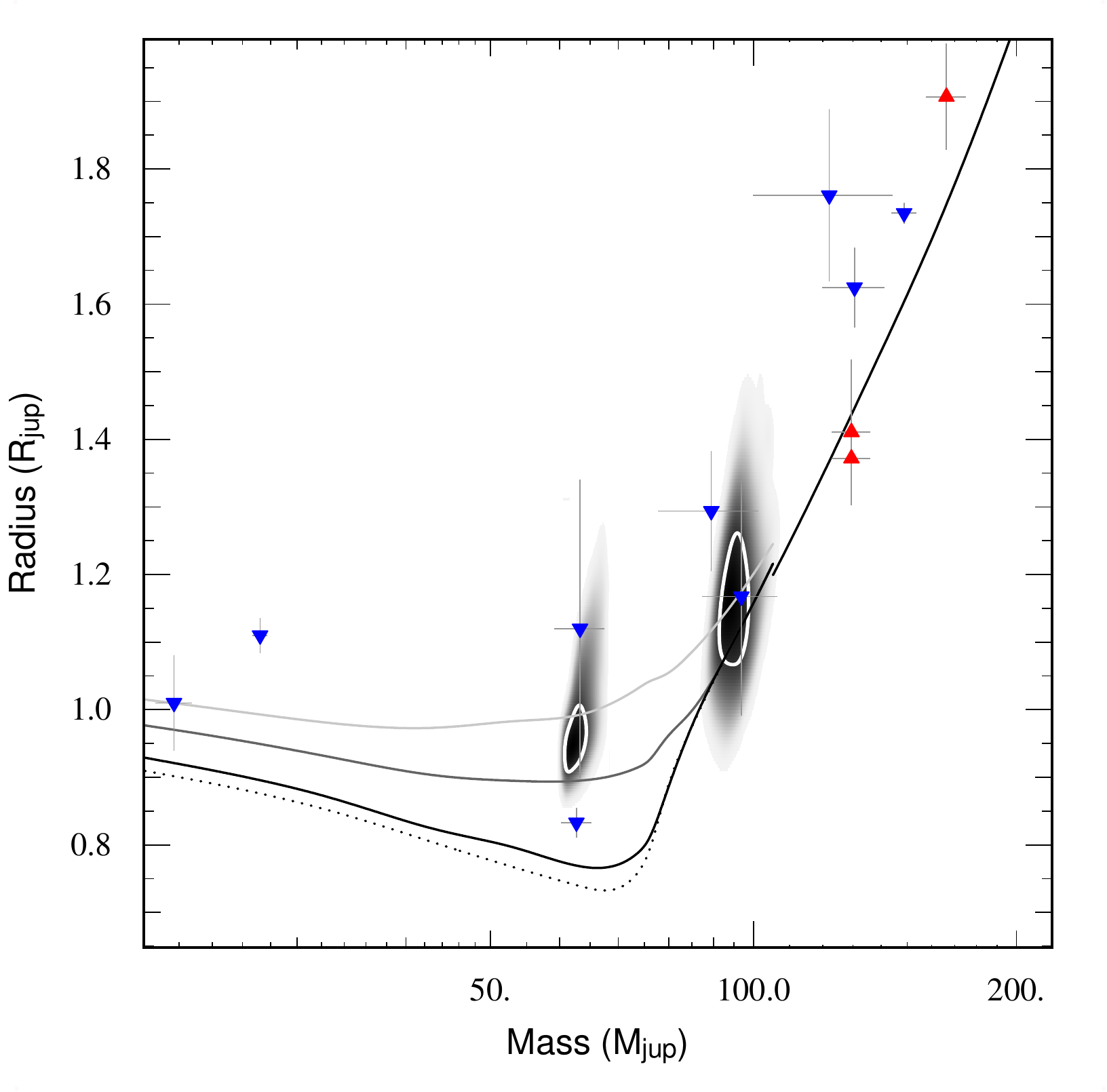}
\caption{Mass--radius diagram for heavy planets, brown dwarfs and low-mass stars. The radius axis corresponds to the size range of Jupiter-mass planets discovered so far. Inverted blue triangles show eclipsing/transiting SB1s, upright red triangles denote interferometric measurements. The two (M$_2$, R$_2$) posterior probability density distributions for WASP-30b and J1219--39b are drawn in grey with their $1-\sigma$ confidence regions in white. Models by \citet{Baraffe:2003gf,Baraffe:1998ly} are also displayed with ages 5\,Gyr (black), 1\,Gyr (dark grey), 500\,Myr (light grey) and 10\,Gyr (dotted). Models are for [M/H] $= 0 $. Observational data were taken from \citet{Lane:2001kx,Segransan:2003fr,Pont:2005qy,Pont:2005fk,Pont:2006lr,Beatty:2007uq,Deleuil:2008lr,Demory:2009ys,Bouchy:2011lr,Johnson:2011fk,Ofir:2012lr,Siverd:2012fk}.}
\label{fig:mr}
\end{figure}

While not being the primary objective of this paper, we present a method to analyse, in a global manner, eclipse photometry, the radial-velocity reflex motion, and the Rossiter--McLaughlin effect, for objects more massive than planets, in order to obtain precise estimates of the mass, radius and orbital parameters of SB1s. The precision of a few percent that we obtain comes from our use of $\rho_\star$, the mean stellar density, instead of the more traditional $\log\,g$ when interpolating inside the stellar evolution tracks. This interpolation gives us precise values for the primary's stellar parameters which are used to estimate the secondary's parameters. We can check our method by deducing $\log\,g_\star$ and comparing them to their spectral counterpart. The values are in very good agreement and fall within the errors of the spectral method (see table \ref{sec:results}). This method also allows us to estimate ages from reading the stellar tracks, something important in the case of close binaries where gyrochronology cannot be trusted owing to the tidal evolution of the system.

Our low-mass eclipsing objects have very similar surface gravities, but, located on opposite sides of the brown-dwarf limit, have sizes dominated by different physics \citep{Baraffe:2003gf}. Plotting the posterior probability distribution for the mass and radius on theoretical mass--radius relations \citep{Baraffe:2003gf,Baraffe:1998ly} in Figure \ref{fig:mr}, we observe that WASP-30b is between the 0.5 and 1 Gyr tracks, suggesting the object is fairly young (and thus luminous, which should cause a measurable secondary eclipse). This is at odds with the age of the primary, which we have found to be older than 2 Gyr with $99\%$ confidence, with a best age of $3.4\pm0.4$ Gyr (figure \ref{fig:w30age}). 

We could explain this radius anomaly if the object has been inflated in the same manner that has been observed for hot Jupiters thanks to the high irradiation received from its primary \citep{Demory:2011lr}. The exact physical causes are still being debated. It may also be that energy has been stored inside the object if it circularised from a previously highly eccentric orbit \citep{Mazeh:1979eu,Fabrycky:2007pd}.  WASP-30b's mass is interesting in that it is close to the minimum of the tracks presented in \citet{Baraffe:2003gf} and displayed in figure \ref{fig:mr}. The mass--radius posterior distribution of J1219--39b is compatible with the 10-Gyr theoretical line. Better photometry would help to reduce the confidence region. This analysis will be the subject of subsequent papers (Hebb et al. in prep). One could object that our analysis does not take into account the fact that both WASP-30b and J1219--39b are self luminous. Even in the case that they had the same effective temperature as their primaries, the overall contamination cannot be larger than their relative sizes $\sim 1\,\%$, lower than our current precision.

We would like to attract the attention on the fact that both those objects have sizes entirely compatible with those of hot Jupiters. While hot Jupiters are often inflated, Jupiter-mass planets at longer periods no longer are \citep{Demory:2011lr}. It would then be expected that  many of the planet candidates published by the space mission \textit{Kepler}  with inflated radii ($> 1.2\,R_\mathrm{Jup}$) and periods longer than 10--15 days could be objects similar to WASP-30b and J1219--39b. 
While not being planets, they are of great interest for their masses, radii and orbital parameters.\\

Finally, the Rossiter--McLaughlin effect has recently been used almost exclusively to measure planetary orbital planes. Observing it for binary stars extends that work by bridging the gap in mass ratio between planetary and stellar systems. Comparison between low-mass binaries in our case with higher-mass binaries as in the BANANA survey \citep{Albrecht:2011bs} will permit us to test different regimes of binary formation and tidal interactions. 

Yet more information still lies in the study of this RV anomaly. Its use in the beginning of the $20^\mathrm{th}$ century was primarily to measure the rotation of stars. As seen in this paper, there is a discrepancy between the \vsini values obtained by using calibration of the macroturbulence and the directly measured $V\,\sin\,i_\star$ from the Rossiter--McLaughlin effect. This was also pointed out in \citet{Triaud:2011vn} in the case of WASP-23, and by \citet{Brown:2012lr} in the case of WASP-16. Both those systems, like J1291--39, contain K dwarfs, and both had their spectroscopic $v\,\sin\,i_\star$ overestimated compared to the value obtained via the Rossiter--McLaughlin effect. Since we compute the spin--orbit angle $\beta$ and the $V\,\sin\,i_\star$, we have strong constraints on the coplanarity of the system and rotation velocity of the primary; thanks to the transit/eclipse geometry, we obtain accurate and precise masses and radii for both the primary and the secondary. Combining all this information and collecting many measurements, we will be able to test which of the macroturbulence laws one should use. Should none apply, inserting the observed $V\,\sin\,i_\star$ values as input parameters in spectral line analyses we will have the capacity to measure macroturbulence directly. Observing the Rossiter--McLaughlin effect is thus not just about glimpsing into the past dynamical history of systems, but can also become an important tool for  understanding  stellar physics better.

\paragraph{\textbf{Nota Bene}}
We used the UTC time standard and Barycentric Julian Dates in our analysis. Our results are based on the equatorial solar and jovian radii and masses taken from Allen's Astrophysical Quantities.

\begin{acknowledgements} 
The authors would like to acknowledge the use of ADS and of \textit{Simbad} at CDS. We also would like to attract attention on the help and kind attention of the ESO staff at La Silla and at the guesthouse in Santiago as well as on the dedication of the many observers whose efforts during many nights were required to obtain all the data presented here. Special thanks go to our programmers and their wonderful automatic Data Reduction Software, allowing us to observe live (!) an object transiting its primary (be it planet, brown dwarf or M dwarf), thus makes observing so much more exciting. We thank Kris He\l miniak for commenting on the draft and generally for being a nice guy. Thanks also go to Brice-Oliver Demory for his help with regards with interferometric masses and radii measurements of low-mass stars and the inspiration he provided to AT for starting to seek brown dwarfs, which led to the discovery of those many low-mass binaries.  This work is supported financially by the Swiss Fond National de Recherche Scientifique. 
\end{acknowledgements}

\bibliographystyle{aa}
\bibliography{../1Mybib.bib}

\appendix

\begin{table*}
\caption{CORALIE radial velocities on WASP-30.  Points excluded from the analysis are marked by \#.}\label{obs:wasp30cor}
\centering
\begin{tabular}{ccccc c}
\hline
\hline
&JDB-2\,400\,000 	& RV			&$1\,\sigma$ error	&bissector span & exposition time\\
&day				& m s$^{-1}$	&m s$^{-1}$&m s$^{-1}$&s\\
\hline

&55009.906531&	14.26750&	0.03161&	0.08307	&1800.687\\
&55040.872183&	1.29532&	0.04933&	-0.18938	&1800.683\\
&55092.697730&	14.34312&	0.04450&	0.04157	&1800.680\\
&55095.689403&	5.16045&	0.02890&	-0.04933	&1800.678\\
&55096.547613&	12.97375&	0.03861&	-0.01846	&1800.680\\
&55096.871175&	14.44650&	0.04125&	0.00031	&1800.680\\
&55097.535138&	12.50573&	0.04066&	0.04896	&1800.680\\
&55097.873525&	9.72874&	0.04880&	0.04062	&1800.680\\
&55098.553830&	3.34309&	0.03992&	-0.00745	&1800.678\\
&55113.520897&	14.44839&	0.03112&	-0.13446	&2700.520\\
&55113.587712&	14.55705&	0.03513&	-0.01233	&1800.676\\
&55113.611079&	14.50745&	0.03642&	-0.01244	&1800.676\\
&55113.634342&	14.57909&	0.03446&	-0.10343	&1800.676\\
&55113.657605&	14.48377&	0.03133&	-0.03359	&1800.676\\
&55113.680879&	14.52860&	0.02992&	-0.06087	&1800.677\\
&55113.704143&	14.50256&	0.03056&	-0.06218	&1800.677\\
&55372.894609&	3.43134&	0.02395&	0.02263	&1800.692\\
&55374.882333&	11.71436&	0.02861&	-0.07726	&1800.692\\
&55375.912731&	13.35947&	0.02948&	-0.00708	&1800.690\\
&55376.943164&	4.24058&	0.02456&	0.02991	&1800.690\\
\#&55406.669984&	21.10817&	0.15149&	-0.69141	&600.570\\
\#&55406.679382&	21.19725&	0.16601&	-0.56823	&600.570\\
\#&55406.688770&	21.08917&	0.17616&	0.44692	&600.570\\
\#&55406.698180&	20.98105&	0.16437&	-0.31414	&600.570\\
\#&55406.707903&	21.09235&	0.16569&	-0.88959	&600.570\\
\#&55406.717302&	21.04424&	0.14780&	0.31012	&600.570\\
\#&55406.726712&	21.04202&	0.19272&	-1.53008	&600.570\\
\#&55406.736354&	3.44740&	0.16206&	4.13793	&600.570\\
\#&55406.745753&	3.99157&	0.17244&	2.36916	&600.570\\
\#&55406.755152&	21.61630&	0.16635&	0.06005	&600.570\\
\#&55406.767595&	1.62258&	0.11359&	-0.28920	&600.570\\
\#&55406.776994&	4.15874&	0.16749&	5.89641	&600.570\\
\#&55406.786520&	21.54117&	0.16369&	-0.26250	&600.570\\
&55483.592040&	14.52330&	0.02797&	0.05140	&1800.686\\
&55484.555214&	8.71115&	0.03915&	0.01610	&1200.133\\
&55484.571637&	8.57906&	0.03658&	-0.04975	&1200.133\\
&55484.591612&	8.35231&	0.02803&	-0.13637	&1800.685\\
&55484.614887&	8.11803&	0.02732&	-0.11433	&1800.685\\
&55484.638161&	7.80482&	0.02456&	0.05655	&1800.685\\
&55484.661528&	7.56473&	0.02306&	0.08690	&1800.686\\
&55484.684814&	7.34139&	0.02417&	-0.07061	&1800.686\\
&55484.708112&	7.13547&	0.02518&	-0.04579	&1800.686\\
&55484.731432&	6.95050&	0.02617&	-0.02408	&1800.686\\
&55484.757021&	6.68697&	0.02934&	0.03535	&1800.686\\
&55485.577214&	1.43623&	0.04333&	0.12378	&900.856\\
&55485.772876&	1.44188&	0.04730&	0.05507	&900.856\\
&55486.755271&	8.38843&	0.04952&	0.00574	&900.856\\
\hline
\end{tabular}
\end{table*}

\begin{table*}
\caption{HARPS radial velocities on WASP-30.}\label{obs:SW1219}
\centering
\begin{tabular}{cccc c}
\hline
\hline
JDB-2\,400\,000 	& RV			&$1\,\sigma$ error	&bissector span & exposition time\\
day				& m s$^{-1}$	&m s$^{-1}$&m s$^{-1}$&s\\
\hline
55458.801056	&14.31817	&0.01359	&-0.02473	&900.001\\
55458.863070	&14.18789	&0.01721	&-0.02845	&900.000\\
55459.584455	&8.92639	&0.01303	&-0.04849	&900.000\\
55459.596585	&8.79993	&0.01243	&-0.04828	&900.001\\
55459.608645	&8.71706	&0.01227	&0.01562	&900.001\\
55459.619941	&8.59953	&0.01115	&-0.03980	&900.001\\
55459.630832	&8.51616	&0.01139	&-0.02751	&900.001\\
55459.641619	&8.39399	&0.01172	&0.03252	&900.001\\
55459.652406	&8.32565	&0.01163	&-0.09640	&900.001\\
55459.663193	&8.19032	&0.01169	&-0.08725	&900.001\\
55459.674084	&8.08172	&0.01227	&-0.07852	&900.001\\
55459.684663	&7.92218	&0.01182	&-0.00659	&900.001\\
55459.695554	&7.81028	&0.01274	&-0.01962	&900.001\\
55459.706341	&7.70502	&0.01233	&0.02461	&900.001\\
55459.717140	&7.57160	&0.01221	&0.07484	&900.000\\
55459.727822	&7.46093	&0.01280	&0.06383	&900.000\\
55459.738714	&7.36897	&0.01172	&0.00101	&900.001\\
55459.749709	&7.26729	&0.01124	&-0.06186	&900.001\\
55459.760183	&7.15544	&0.01068	&-0.00859	&900.000\\
55459.771086	&7.09373	&0.01130	&0.02301	&900.001\\
55459.781769	&6.97189	&0.01063	&-0.01718	&900.001\\
55459.792938	&6.87136	&0.01065	&-0.00295	&900.001\\
55459.803516	&6.76433	&0.01139	&0.03844	&900.001\\
55459.814199	&6.65709	&0.01154	&0.00489	&900.001\\
55459.824986	&6.53058	&0.01294	&0.06997	&900.001\\
55459.836190	&6.42898	&0.01362	&-0.01800	&900.000\\
55459.846352	&6.34038	&0.01442	&0.06821	&900.001\\
55459.857555	&6.18629	&0.01672	&0.11131	&900.001\\
55460.555502	&0.55393	&0.02949	&1.38692	&900.000\\
55462.734059	&14.44197	&0.01434	&-0.08221	&900.000\\
55462.837403	&14.44747	&0.01748	&-0.08063	&900.001\\
55463.575646	&10.48865	&0.01369	&-0.14413	&900.001\\
55464.554164	&2.07849	&0.01635	&0.08127	&900.001\\
55464.754091	&1.42238	&0.01348	&-0.01207	&900.000\\
55464.821243	&1.20493	&0.02133	&0.20388	&900.001\\
55465.751553	&6.19770	&0.01200	&0.00173	&1200.001\\
55465.836795	&6.99038	&0.02404	&-0.22238	&900.001\\

\hline
\end{tabular}
\end{table*}

\begin{table*}
\caption{CORALIE radial velocities on J1219-39.  Points excluded from the analysis are marked by \#.}\label{obs:SW1219}
\centering
\begin{tabular}{ccccc c}
\hline
\hline
&JDB-2\,400\,000 	& RV			&$1\,\sigma$ error	&bissector span & exposition time\\
&day				& m s$^{-1}$	&m s$^{-1}$&m s$^{-1}$&s\\
\hline
&54682.496675&	23.53814&	0.00542&	-0.02097&	1800.678\\
&54815.860532&	35.05819&	0.00643&	-0.02585&	1500.417\\
&54816.853869&	26.13893&	0.00594&	-0.03457&	1800.677\\
&55041.524239&	25.66908&	0.00662&	-0.01387&	1800.682\\
&55241.863326&	33.73800&	0.00602&	0.00296&	900.845\\
\#&55299.811179&	37.46028&	0.02509&	0.04919&	1800.689\\
&55300.613772&	43.84828&	0.00792&	-0.03227&	900.857\\
&55301.784939&	42.61375&	0.00748&	-0.01918&	900.858\\
\#&55302.558918&	35.39085&	0.07415&	0.00307&	900.859\\
&55305.684930&	29.22708&	0.00654&	-0.01949&	900.859\\
&55308.790500&	40.60277&	0.00877&	-0.03277&	600.571\\
&55310.651007&	24.53962&	0.00740&	-0.01631&	600.570\\
&55329.678564&	34.46613&	0.01404&	0.02957&	600.570\\
\#&55329.687962&	34.41273&	0.02722&	0.00544&	600.570\\
&55329.720217&	33.98013&	0.00859&	0.00912&	900.857\\
&55329.732913&	33.86562&	0.00933&	-0.00727&	900.857\\
&55329.744070&	33.75440&	0.01145&	0.00512&	600.570\\
&55329.755690&	33.62545&	0.01319&	-0.00539&	600.570\\
\#&55329.765239&	33.52162&	0.02340&	-0.00788&	600.570\\
&55329.774717&	33.44389&	0.01553&	-0.07355&	600.570\\
&55390.460520&	35.06057&	0.00926&	-0.01522&	600.570\\
&55390.469906&	34.95219&	0.00989&	-0.00022&	600.570\\
&55390.479280&	34.84649&	0.01008&	-0.03039&	600.570\\
&55390.488654&	34.78109&	0.01226&	-0.02240&	600.570\\
\#&55390.498133&	34.71316&	0.02654&	-0.04848&	600.570\\
\#&55390.507553&	34.55764&	0.05742&	-0.07601&	600.570\\
\#&55390.516939&	34.51794&	0.04491&	-0.25450&	600.571\\
\#&55390.526325&	34.39778&	0.03611&	-0.05093&	600.571\\
\#&55390.535699&	34.24055&	0.02948&	0.01679&	600.570\\
\#&55390.545073&	34.14585&	0.02327&	-0.00550&	600.571\\
&55390.554725&	34.04639&	0.01295&	-0.05073&	600.570\\
&55390.564100&	33.96213&	0.01171&	-0.05840&	600.570\\
&55390.573474&	33.85961&	0.01183&	-0.03258&	600.570\\
&55390.582952&	33.78040&	0.01189&	-0.02163&	600.570\\
&55390.592500&	33.67088&	0.01068&	-0.02565&	600.570\\
&55390.601863&	33.56572&	0.01378&	-0.06646&	600.570\\
&55390.611249&	33.48312&	0.01541&	-0.03775&	600.570\\
&55391.523964&	25.74355&	0.01168&	0.00418&	600.570\\
&55629.824349&	26.70333&	0.01020&	-0.00577&	600.582\\
&55635.796335&	23.62243&	0.01281&	0.00327&	600.580\\
&55644.691444&	38.61303&	0.00849&	0.00907&	600.580\\
&55646.783167&	40.68356&	0.00941&	-0.02779&	600.580\\
&55666.518971&	44.39380&	0.01250&	0.01109&	600.661\\
&55667.592221&	35.34860&	0.00960&	-0.00970&	600.581\\
&55667.601631&	35.25073&	0.00937&	-0.04526&	600.581\\
&55667.611017&	35.16217&	0.00950&	-0.04016&	600.581\\
&55667.620404&	35.07056&	0.00973&	-0.03097&	600.581\\
&55667.629790&	34.96315&	0.00926&	-0.01371&	600.581\\
&55667.639292&	34.86310&	0.00890&	-0.00189&	600.601\\
&55667.648795&	34.78823&	0.00923&	-0.03870&	600.581\\
&55667.658181&	34.67728&	0.00918&	-0.01658&	600.642\\
&55667.667626&	34.58564&	0.00859&	-0.00233&	600.581\\
&55667.677163&	34.48485&	0.00882&	-0.00674&	600.662\\
&55667.686561&	34.38737&	0.00882&	-0.01188&	600.581\\
&55667.695936&	34.25795&	0.00907&	-0.00941&	600.581\\
&55667.705322&	34.13713&	0.00974&	-0.04686&	600.581\\
&55667.714720&	34.05206&	0.00972&	-0.02112&	600.581\\
&55667.724107&	33.93877&	0.00968&	-0.00701&	600.641\\
&55667.733494&	33.87283&	0.00979&	-0.03358&	600.581\\
&55667.742880&	33.76460&	0.00968&	-0.01320&	600.641\\
&55667.752313&	33.68215&	0.00910&	0.01805&	600.581\\
&55667.761699&	33.58665&	0.00940&	0.01132&	600.621\\
&55667.773401&	33.46342&	0.01030&	-0.02990&	601.570\\
&55667.782799&	33.33829&	0.00983&	-0.05807&	601.671\\
&55667.794419&	33.25856&	0.01239&	-0.00666&	601.570\\
&55667.804060&	33.14646&	0.01293&	-0.02891&	601.691\\
&55667.813459&	33.04706&	0.01229&	-0.00366&	601.570\\
&55667.824049&	32.95409&	0.00980&	-0.01739&	800.795\\
&55667.835924&	32.81781&	0.00934&	-0.01553&	800.755\\
&55667.847648&	32.70209&	0.00989&	-0.03184&	800.836\\
&55668.611754&	26.18419&	0.00668&	-0.00885&	800.755\\

\hline
\end{tabular}
\end{table*}

\end{document}